\documentclass[11pt]{articleKI}

\title{\bf Linking tracklets over the years in large datasets}

\author[1,2]{\'Oscar Rodr\'iguez\,\orcidlink{0000-0002-4545-5135}\thanks{\href{mailto:oscar.rodriguez@upc.edu}{oscar.rodriguez@upc.edu}}}
\author[3]{Giovanni F. Gronchi\,\orcidlink{0000-0003-1294-0633}}
\author[3]{Giulio Ba\`u\,\orcidlink{0000-0002-9857-0866}}
\author[4]{Robert Jedicke\,\orcidlink{0000-0001-7830-028X}}

\affil[1]{Dept. Matemàtiques, Universitat Politècnica de Catalunya, Av Diagonal 647, 08028 Barcelona, Spain}
\affil[2]{IMTech, Universitat Politècnica de Catalunya, Pau Gargallo 14, 08028 Barcelona, Spain}
\affil[3]{Dipartimento di Matematica, Universit\`a di Pisa, Largo B. Pontecorvo 5, 56127 Pisa, Italy}
\affil[4]{Institute for Astronomy, University of Hawai`i,
2680 Woodlawn Drive, Honolulu, HI 96822-1839 
USA
}

\newcommand{\au}{\,\mathrm{au}}
\newcommand{\km}{\,\mathrm{km}}
\newcommand{\rps}{\ensuremath{r_{\rm P1}}}
\newcommand{\ips}{\ensuremath{i_{\rm P1}}}

\begin{document}

\maketitle

\begin{abstract}
    We present a new procedure to identify observations of known
    objects in large data sets of unlinked detections.  It begins with
    a Keplerian integrals method that allows us to link two tracklets,
    computing preliminary orbits, even when the tracklets are
    separated in time by a few years. In the second step, we
    represent the results in a `graph' where the tracklets are the
    nodes and the preliminary orbits are the edges. Then, acceptable
    `3-cycles' are identified and a least squares orbit is computed
    for each of them.  Finally, we construct sequences of $n \geq 4$
    tracklets by searching through the orbits of nearby 3-cycles and
    attempting to attribute the remaining tracklets.  We calculate the
    technique's efficiency at identifying unknown objects using real
    detections that attempt to mimic key parameters of the Minor
    Planet Center's Isolated Tracklet File (ITF) and then apply the
    procedure to the ITF to identify tens of thousands of new
    objects.\\[1.5ex]

    \noindent\textbf{Keywords:} Orbit determination, Keplerian
    integrals methods, Linkage problem, Asteroid surveys.
\end{abstract}

\section{Introduction}

In recent years, there have been significant developments in the
observational techniques employed for detecting asteroids which have
resulted in a marked increase in the number of asteroid detections.
This trend is anticipated to continue with the forthcoming surveys,
such as the Vera C. Rubin Observatory's Large Synoptic Survey
Telescope (VRO-LSST) \cite{lsst}, which will survey the sky more
comprehensively and deeply than previous endeavors.  The VRO-LSST is
expected to detect millions of asteroids, including many that are too
small or too faint to be detected by current surveys. This will
provide a wealth of data and will help us to better understand the
population of asteroids in our solar system.

Detections of asteroids are usually grouped into tracklets of very
short arcs (VSA), each referring to the same observed object.  These
tracklets are collected over a few days and are used to compute the
orbit of an asteroid.  If the tracklets are not successfully used to
compute an orbit they are stored in the isolated tracklet file (ITF)
\cite{itflink}, a database maintained by the Minor Planet Center. The
data in the ITF are mainly provided by the Pan-STARRS1
\citep{Denneau2013-MOPS} and Catalina surveys \citep{Catalina} which
are both large programs that have been successful in discovering and
tracking asteroids.  These two observatories have provided more than 4
and 2 million observations, respectively.  With the work done in
recent years, see e.g. \citep{Sansaturio2012,HelioLinC2018,
  WerykITF2020}, the size of the ITF has been considerably reduced.
\citep{Sansaturio2012} used an identification technique called
"attribution type" to compute the orbits of asteroids, particularly
near-Earth asteroids (NEAs), by taking into account their higher
apparent rates of motion.  HelioLinC \citep{HelioLinC2018} used a
tracklet clustering technique to define an algorithm with a complexity
of $\mathcal{O}(N\mathrm{log}N)$, where $N$ is the total number of
tracklets.  \citep{WerykITF2020} developed techniques to optimize the
multi-apparition linking of tracklets based on their apparent rates of
motion despite being far from their predicted locations.

Most initial orbit determination methods
\citep{laplace,lagrange,gauss1809,moss2021} are based on the two-body
equations of motion and rely on Taylor's series expansions around a
central time.  If the detections are widely spaced in time the initial
orbit may not be accurate or may not be computable.


The Keplerian integrals (KI) methods \citep{gbm15,gbm17} impose the
conservation laws of Kepler's dynamics (angular momentum, Laplace-Lenz
vector, and energy) to calculate a preliminary orbit from the
information contained in two or three tracklets.  The main advantage
of these methods is that they do not impose constraints on the time
separation between the tracklets.  The idea of using conservation laws
was introduced by Taff and Hall, who used the conservation of angular
momentum and energy to solve the problem of linking tracklets and
computing preliminary orbits (see for example \cite{TaffHall1977,
  Taff1984}) but did not fully exploit the algebraic character of the
resulting equations, even if they observed that the equations could be
expressed in polynomial form. In these references the high sensitivity
of the equations to astrometric error was already noted.  Later,
\cite{gdm10,gfd11,gbm15} derived polynomial equations of degree 48,
20, and 9, respectively, from the Keplerian conservation laws for the
purpose of linking two VSAs.  Then, \cite{gbm17} demonstrated that the
polynomial of degree 9 introduced in \cite{gbm15} is optimal in some
sense and derived an equation of degree 8 for the linkage of three
VSAs.  \cite{Link2Link3} examined the numerical behavior of two
Keplerian integral algorithms introduced in \cite{gbm15} and
\cite{gbm17}, referred to as \texttt{link2} and \texttt{link3},
respectively.  Although these methods are sensitive to astrometric
error, their analysis showed that solutions with moderate error are
promising.  In addition to their ability to link tracklets that are
widely spaced in time, these methods have the advantage of being
computationally efficient due to their polynomial formulation.

In this study, we propose a procedure for computing least squares (LS)
orbits using ITF detections submitted by Pan-STARRS1 (hereafter
denoted by its observatory code F51) which is known for its small
astrometric errors \cite{Carpino2003}.  The procedure first links
pairs of tracklets using the \texttt{link2} algorithm, then constructs
`3-cycles' composed of 3 tracklets that have been successfully linked
in the previous step.  For each 3-cycle, a `norm' is calculated with
all the orbits obtained by \texttt{link2} using the 3 pairs of
tracklets within the 3-cycle, and only 3-cycles with a norm below a
certain threshold are retained.  For each accepted 3-cycle, we compute
a least squares orbit along with its root mean square (rms)
astrometric error.  Finally, we construct `$n$-ids', sequences of
$n\ge 4$ tracklets that were successfully linked by \texttt{link2}, by
identifying additional candidate tracklets to the 3-cycles and
applying differential corrections.



This article is structured as follows.  In Section~\ref{s:procedure}
we present the proposed procedure for linking tracklets, including the
relevant indicators to assess the quality of the results.
In Section~\ref{s:testproc} the values of the indicators are tuned to
optimize the performance of the algorithm by applying it to a test
data set constructed from real observations of main belt asteroids and
some NEAs.
Finally, in Section~\ref{s:itfproc}, the procedure is applied to all
the F51 tracklets contained in the ITF with at least 3 observations
each.

\section{The procedure}
\label{s:procedure}

Let us consider a list $T$ of $N$ tracklets, each composed of at least
3 observations, with the goal of identifying all the tracklets that
belong to the same objects and determining their orbits.  Our
procedure follows three major steps described in the next
sub-sections.

\subsection{First step: \texttt{link2} exploration}
\label{s:link2all}

The first step is to attempt to link all possible pairs of tracklets
in the list $T = \{t_1,...,t_N\}$.  Since this step has a quadratic
cost, i.e. $\mathcal{O}(N^2)$, a highly efficient method is necessary
if the number of tracklets $N$ is large, as in the case of the ITF.
Additionally, the linking method must be able to join tracklets even
if they are separated by several years.

The KI algorithm \texttt{link2}, introduced in \cite{gbm15} and tested
in \cite{Link2Link3}, is well-suited for this purpose.  It is based on
solving a univariate polynomial of degree 9 in the radial distance of
the observed object at the mean epoch of one of the two tracklets
being linked.  The use of a polynomial with a relatively low degree
makes this method fast compared to others.  The \texttt{link2}
algorithm is not symmetric, i.e. it is sensitive to inverting the
order of the two tracklets, so both options must be considered to
ensure that all possible linkages are computed.

In the following, we refer to a \emph{linkage} between two tracklets
as a successful join using the \texttt{link2} algorithm, without any
quality control.  However, when using a KI method in practice,
it is possible to obtain a preliminary orbit even if the tracklets do
not belong to the same object, and to obtain a poor quality
preliminary orbit even if they do.  It is desirable to maximize the
number of \emph{true} linkages (where the tracklets belong to the same
object), while minimizing the number of \emph{false} ones.

The $\chi_4$ and $rms$ metrics \cite{Link2Link3} are useful to
quantify the quality of the solutions and select the best ones for the
next step.  Thresholds for $\chi_4$ and $rms$ were determined based on
testing with real observations of known objects
(Section~\ref{s:testproc}).  \texttt{link2} linkages that satisfy the
threshold values of both metrics are referred to as \emph{accepted}.

Since it is possible to obtain more than one solution for each pair of
tracklets, as we are computing roots of polynomials of degree 9, we
will denote by $o_{ij}^{(k)}$ the $k$-th accepted preliminary orbit
obtained from tracklets $t_i$, $t_j\in T$ using the \texttt{link2}
method.

Even after filtering solutions with the $\chi_4$ and $rms$ metrics it
is possible to obtain false accepted linkages, i.e. accepted linkages
between tracklets that do not belong to the same object.
Additionally, even in the case of true linkages, the computed orbits
are often not sufficiently accurate.  To address these issues, the
next two steps of the procedure are applied to join together more than
two tracklets.

\subsection{Second step: constructing LS orbits using 3 tracklets}

The next step is to group sets of three tracklets $\{t_i,t_j,t_k\}$
using the information obtained from \texttt{link2} such that each pair
within the set is an accepted linkage.

\subsubsection{Constructing 3-cycles}

The results obtained using the \texttt{link2} algorithm can be
represented as a graph $G=G(V,E)$, where the set of vertices $V=
\{1,2,...,N\}$ corresponds to the set of tracklets
$T=\{t_1,...,t_N\}$, and the set of edges $E$ corresponds to the
accepted linkages.  Specifically, $e_{ij}\in E$, with $i,j\in V$,
$i>j$, if and only if a linkage between tracklets $t_i$ and $t_j$ is
found with at least one ordering of the tracklets and with acceptable
values of $\chi_4$ and $rms$.  Therefore, each edge in $E$ represents
an accepted linkage.  Even if there is more than one solution for a
pair of tracklets, $t_i$ and $t_j$, we consider only one edge for the
linkage, i.e. we consider a \emph{simple} graph.  Moreover, we treat
the edges as having a direction, from $i$ to $j$, so it is possible
interpret the graph as a directed graph.  The ordering is introduced
only to simplify the computations.

As previously mentioned, our goal is to search for sets of three
tracklets $\{t_i,t_j,t_k\}$ such that each pair of tracklets is an
accepted linkage.  This is equivalent to searching for sets of
vertices $i,j,k\in V$ such that $e_{ij},e_{ik},e_{jk}\in E$ or, in
other words, searching for 3-cycles in the graph $G$.

\begin{figure}[!ht]
  \centering
  \includegraphics[width=0.6\textwidth]{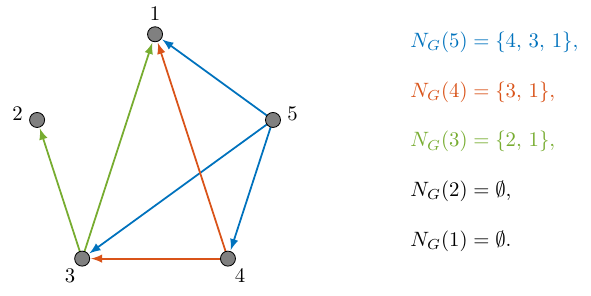}
  \caption{Example of graph and sets $N_G(i)$.}
  \label{fig:graf_exemple}
\end{figure}

To find all the 3-cycles, for each vertex we select all the adjacent
vertices in descending order $i = N,N-1,...,2$, i.e. the set
\[
N_G(i) = \{l\in V\,|\, e_{il}\in E,\, l<i\}.
\]
In addition, the elements of the set $N_G(i)$ are considered in
descending order, that is $N_G(i)=\{l^{(i)}_1,l^{(i)}_2,\ldots
,l^{(i)}_{p_i}\}$, with $l^{(i)}_1 >l^{(i)}_2>...>l^{(i)}_{p_i}$ (see
Figure \ref{fig:graf_exemple}).  Using an adjacency list can save a
significant amount of space compared to other graph representations,
such as an adjacency matrix. The reduction in the space required to
store the graph is especially important when dealing with sparse
graphs, as is our case. It is also easy to insert or delete elements
in the linked list.

\begin{figure}[!ht]
  \centering
  \includegraphics[width=1\textwidth]{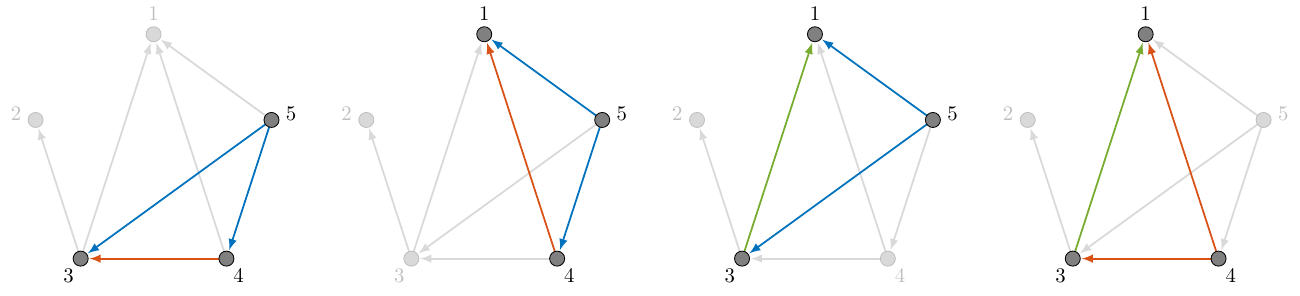}
  \caption{The 3-cycles of the graph of Figure \ref{fig:graf_exemple}.}
  \label{fig:graf_exemple_3cycles}
\end{figure}

Finally, this representation is useful to find all the 3-cycles by the
following procedure: for each $i = N,N-1,...,3$, we find the set of
neighbors $N_G(i)$ and for each $j \in
N_G(i)\setminus\{l^{(i)}_{p_i}\} = \{l^{(i)}_1,...,l^{(i)}_{p_i-1}\}$,
we consider the set of neighbors $N_G(j)$. We then search for indices
$k\neq i,j$ such that $k\in N_G(i) \cap N_G(j)$.  For each tracklet
$k$ that satisfies this condition we obtain the 3-cycle $\{i,j,k\}$
(see Figure \ref{fig:graf_exemple_3cycles} where we display the
3-cycles of Figure \ref{fig:graf_exemple}).  This classical procedure
for searching 3-cycles is detailed in Algorithm \ref{al:3cycles}.

\begin{algorithm}
  \caption{Finding 3-cycles}
  \label{al:3cycles}
  \begin{algorithmic}[1]
    \State $V = \{1,...,N\}$
    \For{$i=N,N-1,...,3$}
    \For{$j \in N_G(i)\setminus\{l^{(i)}_{p_i}\}$}
    \State Save the sets $\{i,j,k\}$ with $k\in N_G(i) \cap N_G(j)$.
    \EndFor
    \EndFor
  \end{algorithmic}
\end{algorithm}

We make the following remarks.

\begin{remark}
  In order to find the 3-cycles in an efficient way it is important
  to sort the set of vertices. This is useful to avoid searching for
  equivalent 3-cycles e.g. $\{i,j,k\}$ and $\{j,i,k\}$.
\end{remark}

\begin{remark}
  Selecting the tracklets in descending order allows us to avoid the
  exploration of the entire graph if we add new tracklets to the
  data set.  In particular, let ${t_{N+1},...,t_{N+M}}$ be the $M$
  new tracklets added to $T$.  The addition of these new tracklets
  corresponds to the inclusion in $G$ of the vertices
  ${N+1,...,N+M}$ and their corresponding edges $e_{ij}$, with
  $i\in\{N+1,...,N+M\}$ and $j\in\{2,...,N+M-1\}$.  To find the new
  3-cycles it is only necessary to perform the first loop of
  Algorithm \ref{al:3cycles} for $i=N+M,...,N+1$.
\end{remark}

\begin{remark}
  Algorithm \ref{al:3cycles} is easily parallelizable by
  distributing the values of $i$ among the different nodes.
\end{remark}

\subsubsection{Angular momentum norm and LS orbits}

For correct linkages \cite{Link2Link3} showed that the angular
momentum of orbits, $\bm{c}$, computed using \texttt{link2} is
accurate.  As a result, we employ the angular momentum as a measure of
the quality of the 3-cycles.  We recall that for each pair of
tracklets \texttt{link2} is applied with both orderings resulting in
possible multiple solutions.  These solutions possess distinct values
of the $\chi_4$ and $rms$ metrics that are taken into account when
evaluating the quality of the 3-cycles.  To quantify the quality of
the 3-cycles we define the {\em angular momentum norm} as follows:
\begin{equation}
  ||\{t_i,t_j,t_k\}||_M = \min_{h, \ell, p}\left\{ m(o_{ij}^{(h)},o_{ik}^{(\ell)}) +
  m(o_{ij}^{(h)},o_{jk}^{(p)}) + m(o_{ik}^{(\ell)},o_{jk}^{(p)})\right\},
  \label{eq:angmomnorm}
\end{equation}
with
\begin{equation*}
  m(o_A,o_B) = \frac{|\bm{c}_A-\bm{c}_B|}{\sqrt{|\bm{c}_A||\bm{c}_B|}}
  \left(\chi_{4,A}+\chi_{4,B}\right)\left(rms_{A}+rms_{B}\right),
\end{equation*}
where the subscripts $A,B$ refer to the orbits $o_A, o_B$. This norm
simply measures the difference between the angular momenta of the
preliminary solutions using the indicators as weights.

After computing norm \eqref{eq:angmomnorm} for all the 3-cycles, we
sort them by its values in ascending order, and accept only the
3-cycles with the norm below a threshold that will be determined
later.

\subsubsection{LS orbits}
\label{s:LSorb}

Finally, for each 3-cycle we construct an orbit by means of the least
squares (LS) method starting with the preliminary orbits from
\texttt{link2} and \texttt{link3}, because
sometimes \texttt{link3} provides a better initial orbit for the
differential corrections than \texttt{link2}.

Not all the 3-cycles will yield a LS orbit because the differential
corrections algorithm may not converge.  Furthermore, a solution will
not be retained if the rms of the residuals of the resulting LS orbit
is not sufficiently small.

We denote by $\mathcal{C}$ the set of triplets of tracklets
$\{t_{i_1},t_{i_2},t_{i_3}\}$ with an acceptable LS orbit $o_i$
ordered by the angular momentum norm. It is important to note that the
same tracklet can be present in multiple 3-cycles.

\begin{algorithm}[!ht]
  \caption{Joining 4 or more tracklets}
  \label{al:4ormore}
  \begin{algorithmic}[1]
    \State $T = \{t_1,...,t_N\}$
    \For{$i=1,...,m$}
    \State $S = \{t_{i_1},t_{i_2},t_{i_3}\}$
    \If{$S \subseteq T$}
    \For{$j=i+1,...,m$}
    \If{($\{t_{j_1},t_{j_2},t_{j_3}\}\subseteq T$) and ($\{t_{j_1},t_{j_2},t_{j_3}\}\nsubseteq S$)
      and ($o_i$, $o_j$ close enough)}
    \State $S^* = S \cup \{t_{j_1},t_{j_2},t_{j_3}\}$
    \State $o \gets \texttt{difCor}(o_i,S^*)$ 
    \If{successful \texttt{difCor}}
    \State $o_i = o $ 
    \State $S = S^*$
    \EndIf
    \EndIf
    \EndFor    
    \For{$t_j\in T\setminus S$}
    \If{$t_j$ is close enough to $o_i$}
    \State $o \gets \texttt{difCor}(o_i,\{S,t_j\})$ 
    \If{successful \texttt{difCor}}
    \State $o_i = o $ 
    \State $S = S \cup \{t_j\}$
    \EndIf  
    \EndIf       
    \EndFor
    
    \If{$\texttt{size}(S)\geq 4$}
    \State Save $S$ and $o_i$.
    \State $T = T\setminus\{S\}$
    \EndIf
    \EndIf
    \EndFor
  \end{algorithmic}
\end{algorithm}

\subsection{Third step: joining 4 or more tracklets}

The last step is to attribute at least one additional tracklet to the
LS orbit. The general idea is that for each triplet in ${\cal C}$ we
attempt to identify other triplets of ${\cal C}$ that have orbits
close to that of the considered triplet, and then try to attribute the
new tracklets to the original triplet.

After applying the second step above we obtained $m$ triplets of
tracklets with a LS orbit, i.e. we have $\mathcal{C} = \{
\mathcal{C}_i = ( \{t_{i_1},t_{i_2},t_{i_3}\},o_i) \text{with}
i=1,...,m \}$. $T$ will denote here the set of tracklets that have not
been assigned to a LS orbit with 4 or more tracklets. Before applying
the third step, $T$ coincides with the set of the $N$ available
tracklets, and as we obtain LS orbits with 4 or more tracklets, these
we will removed from $T$.


We select the elements $\mathcal{C}_i$ of $\mathcal{C}$ following the
order in which they appear in $\mathcal{C}$ and, if the three
tracklets are in $T$, we consider the set of tracklets
$S=\{t_{i_1},t_{i_2},t_{i_3}\}$ and the orbit $o_i$.  Then, for each
element $\mathcal{C}_j\in\mathcal{C}$ with $j=i+1,...,m$ we check if
all tracklets $t_{j_1},t_{j_2},t_{j_3}\in T$ and
$\{t_{j_1},t_{j_2},t_{j_3}\}\nsubseteq S$. If both these conditions
are satisfied, we check whether the orbit $o_j$ is \emph{close enough}
to the orbit $o_i$.  We say that the orbits $o_B$ and $o_A$ are close
enough if their orbital elements satisfy
\[
\left|\frac{a_A-a_B}{a_A}\right|<\varepsilon_a,\quad
|e_A-e_B|<\varepsilon_e,\quad
|i_A-i_B|<\varepsilon_i,\quad
|\Omega_A-\Omega_B|<\varepsilon_\Omega,\quad
|\omega_A-\omega_B|<\varepsilon_\omega,
\]
for some 
sufficiently small values of the $\varepsilon$ thresholds, and if the
tracklets belonging to the orbit $o_B$ and not to $o_A$ are close to
the ones simulated from the orbit $o_A$.\footnote{The latter condition
  is checked by comparing the values of the detections
  $(\alpha_i,\delta_i)$ at epochs $t_i$ of the tracklets related to
  $o_B$ with the simulated detections obtained by propagation of the
  orbit $o_A$ at the same epochs $t_i$.} If the two orbits are close
enough we try to calculate a LS orbit for the detections contained in
the tracklets in $S$ and the new detections using the orbit $o_i$ as
the initial guess. If the differential corrections converge, the
orbit $o_i$ is updated with the LS orbit and the new tracklets are
added to $S$.  The differential corrections are successful if a
solution is obtained using all the observations and the rms of the
residuals of the LS orbit is below a certain threshold.

The second part of the third step consists in trying to attribute the
tracklets remaining in $T$
to the LS orbits that have already been computed. It is not feasible
to apply it to each possible attribution because the differential
corrections algorithm is computationally expensive.
To minimize the number of candidates for a given LS orbit we employ a
criterion based on the attributables.
From the orbit we can compute a propagated attributable $(\alpha_p,
\delta_p, \dot{\alpha}_p, \dot{\delta}_p)$ at the epoch of the
attributable $(\alpha,\delta,\dot{\alpha},\dot{\delta})$ associated
with a tracklet.  We quantify whether the difference between the two
attributables is sufficiently small by requiring
\[
\begin{aligned}   
  &\left(\cos\alpha\cos\delta-\cos\alpha_p\cos\delta_p\right)^2 + 
  \left(\sin\alpha\cos\delta-\sin\alpha_p\cos\delta_p\right)^2 +
  \left(\sin\delta-\sin\delta_p\right)^2 < \varepsilon_1,\\
  &\frac{(\dot{\alpha}-\dot{\alpha}_p)^2+(\dot{\delta}-\dot{\delta}_p)^2}
       {\sqrt{(\dot{\alpha}^2+\dot{\delta}^2)(\dot{\alpha}_p^2+\dot{\delta}_p^2)}} < \varepsilon_2,
\end{aligned}
\]
for some small values of $\varepsilon_1, \varepsilon_2$ and then
proceed with the attribution.

If the differential corrections are successful we update $o_i$ and add
the new tracklet to $S$. It is worth noting that the tracklets in $T$
can be ordered chronologically to reduce the computational cost of the
propagation so that we can reduce propagation times by using the
results of previous propagations.

Finally, if $S$ contains 4 or more tracklets, we save the orbit $o_i$
and the set $S$ of tracklets that were used to compute it and remove
the tracklets in $S$ from $T$.  The schematic idea of the procedure is
described in Algorithm~\ref{al:4ormore}.

Algorithm~\ref{al:4ormore} can be implemented as a sequence of two
separate steps to facilitate its parallelization.
In the first part, we search for the 3-cycles whose orbits are close
enough and compute a least squares orbit. Moreover, we remove the
tracklets from the leftover database $T$ when they have been used to
construct an orbit with 4 or more tracklets.  In the second part, we
try to attribute the tracklets in $T$ to the orbits obtained in the
previous step.
It might happen that in the first step of the algorithm the same
tracklet is employed in the computation of different orbits. In these
cases we keep the orbit obtained from the 3-cycle with the smallest
value of the $M$ norm; the common tracklet is removed from the set of
tracklets of the other orbit(s) and if at least 4 tracklets remain we
try to compute a new least squares orbit, otherwise the tracklets are
added to $T$.

Note that at the end of the first part of the algorithm we may have
orbits obtained from only 3 tracklets. These orbits will be discarded
if at least one additional tracklet is not attributed to them in the
second part.

Finally, when parallelizing the second part of the procedure we may
also obtain inconsistencies due to tracklets that have been attributed
to more than one orbit.  However, we can easily eliminate these
inconsistencies a posteriori.

\section{Testing the procedure}
\label{s:testproc}

To define the values of the thresholds of the norms described in
Section~\ref{s:procedure} we apply the procedure to a set of real F51
observations of known asteroids.


\begin{figure}[!ht]
  \centering
  \begin{subfigure}{0.47\textwidth}
    \includegraphics[width=\linewidth]{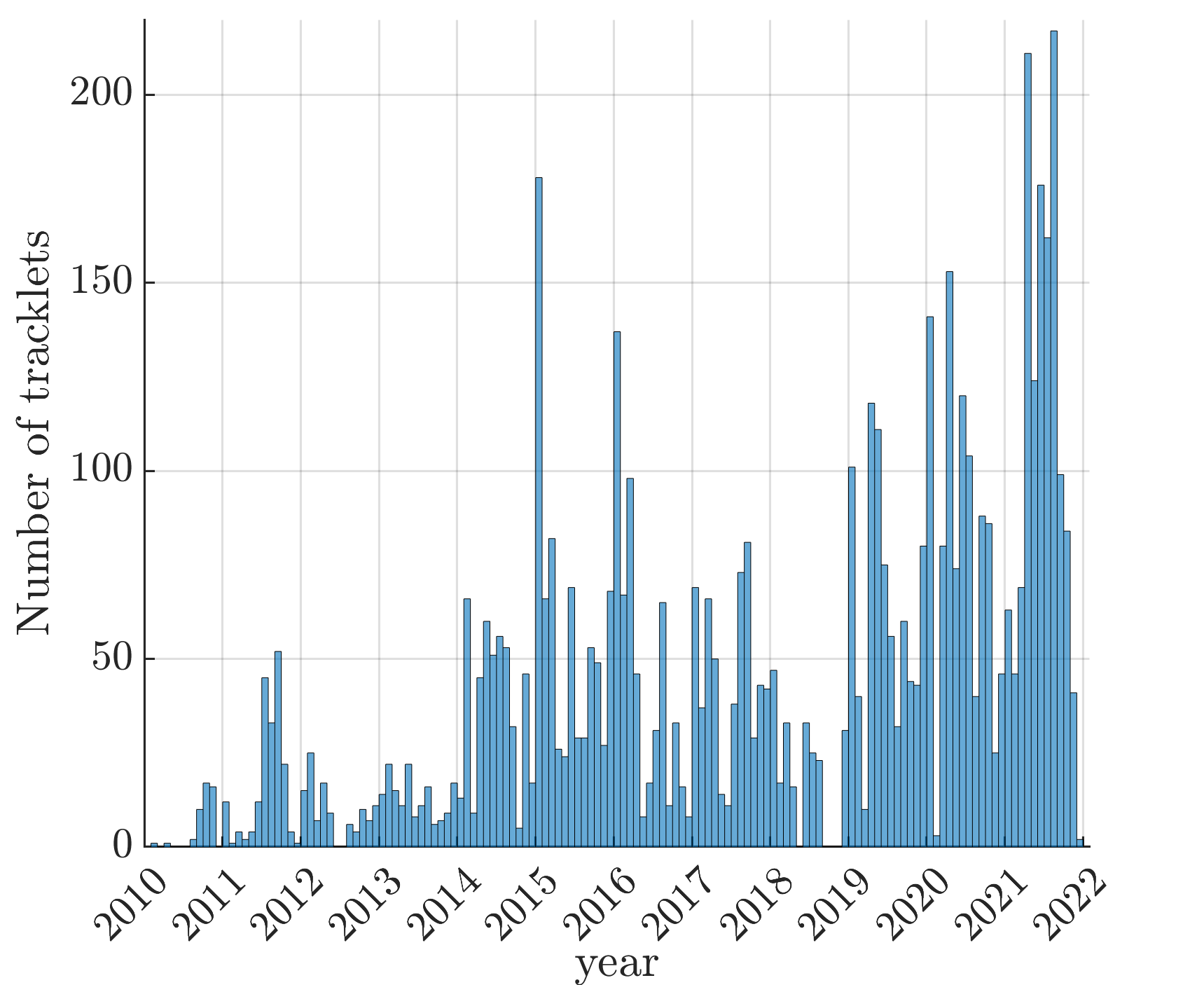}
    \caption{}
    \label{fig:tracklet_time_mag_distn_a}
  \end{subfigure}%
  \begin{subfigure}{0.47\textwidth}
    \includegraphics[width=\linewidth]{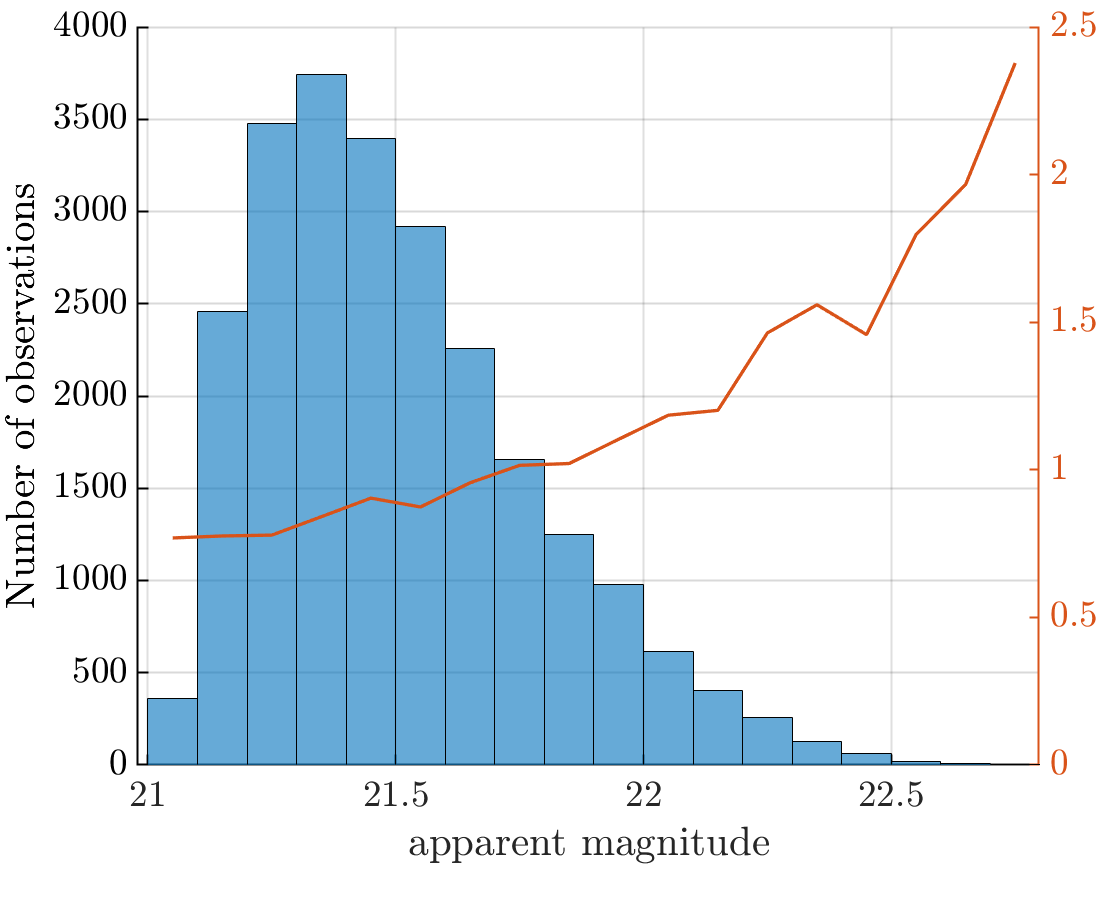}
    \caption{}
    \label{fig:tracklet_time_mag_distn_b}
  \end{subfigure}\\
  \begin{subfigure}{0.47\textwidth}
    \includegraphics[width=\linewidth]{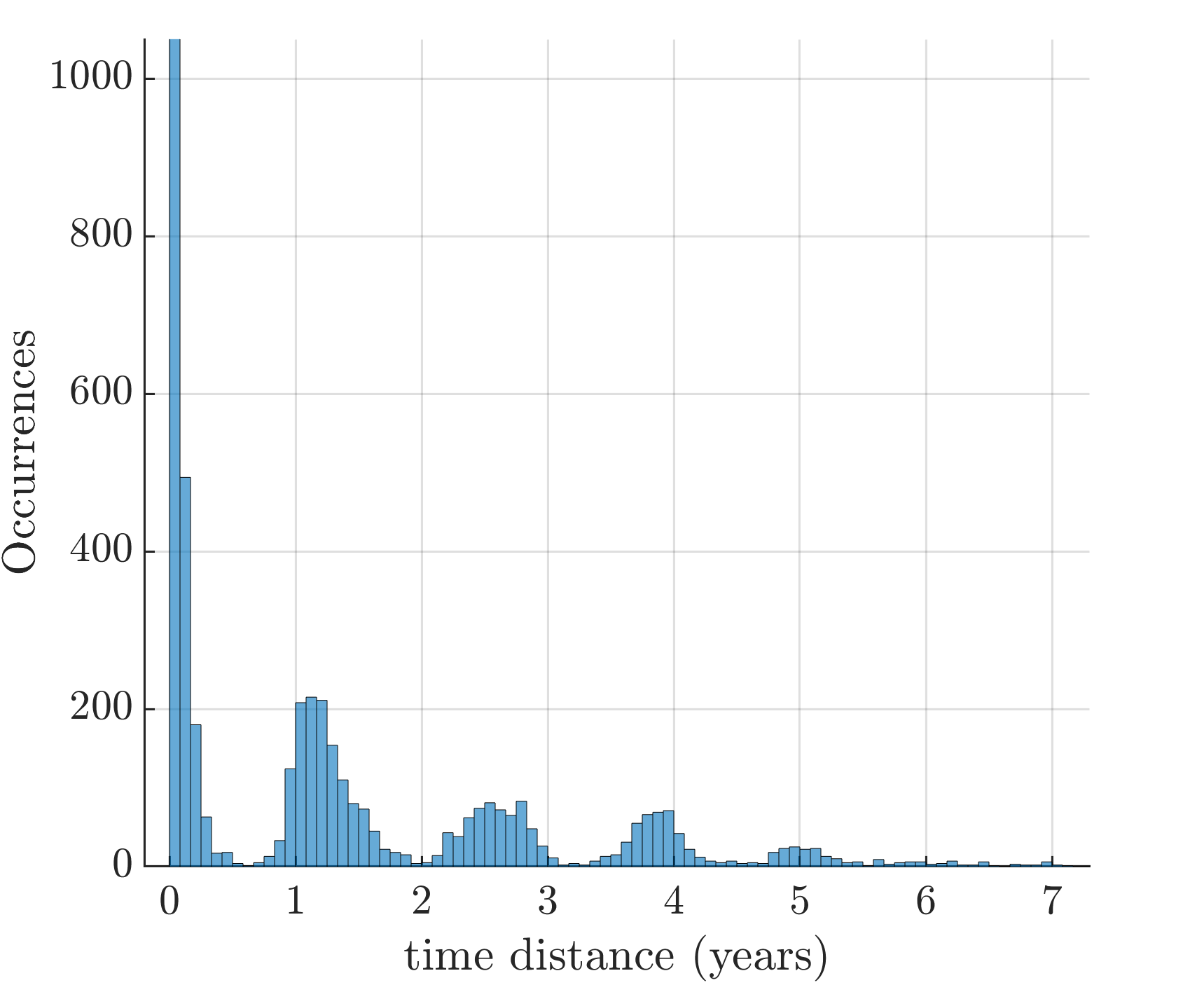}
    \caption{}
    \label{fig:tracklet_time_mag_distn_c}
  \end{subfigure}%
  \begin{subfigure}{0.47\textwidth}
    \includegraphics[width=\linewidth]{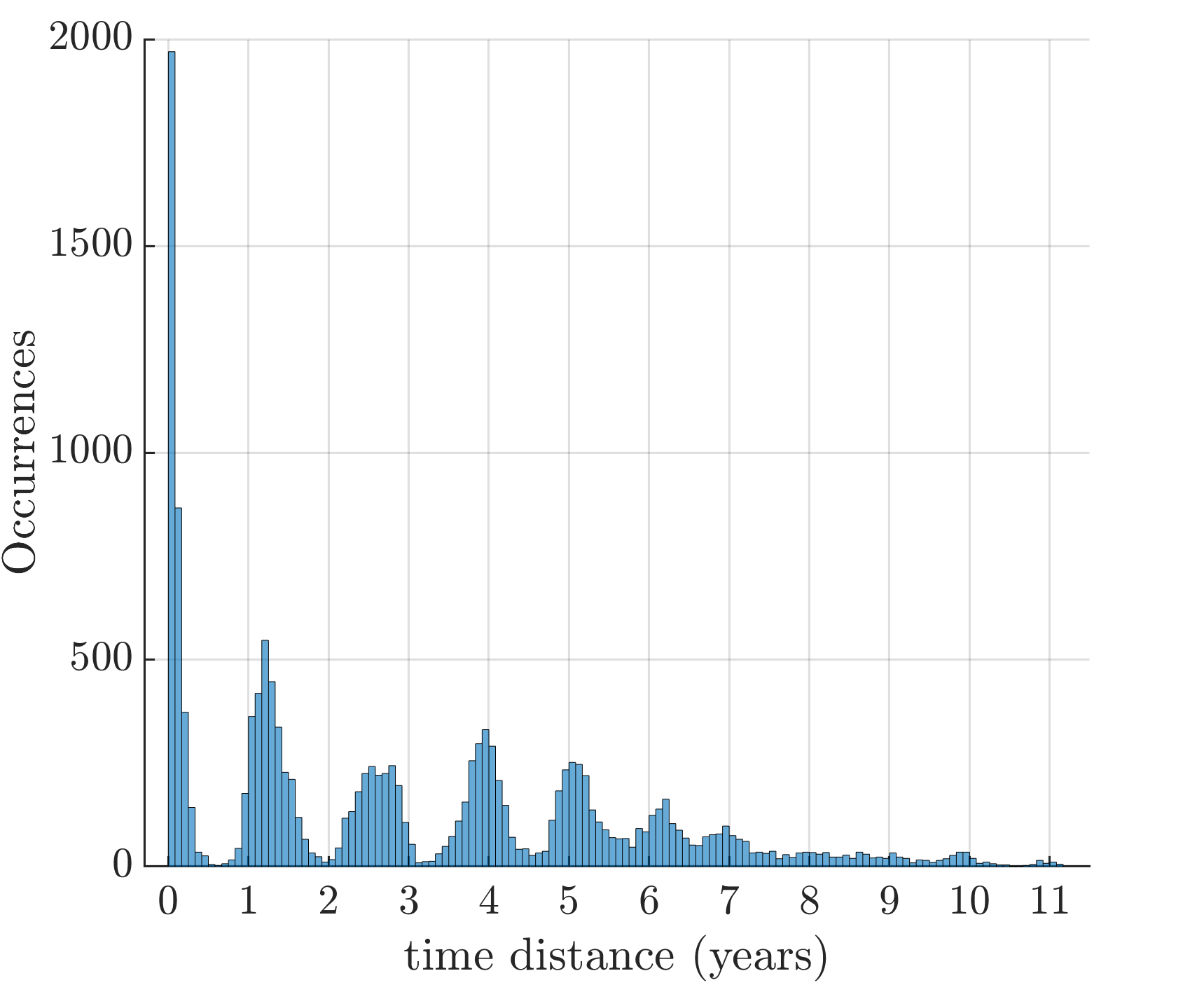}
    \caption{}
    \label{fig:tracklet_time_mag_distn_d}
  \end{subfigure}%
  \caption{
    a) The time of observation of all 6 tracklets for the test data
    sample,
    b) their reported apparent magnitude distribution and (in red)
    average astrometric error,
    c) the time between the closest two pairs of tracklets for each
    object, and
    d) the time between all pairs of tracklets for each object.}
  \label{fig:tracklet_time_mag_distn}
\end{figure}

\subsection{The test dataset}

To test 
our linking algorithm and determine the thresholds for the ITF
processing we extracted a realistic set of ITF-like tracklets from
actual F51 observations.

The test data is composed of real F51 observations of 1021 asteroids
with $\ge6$ tracklets each, where each tracklet contains $\ge4$
detections acquired between 2010 and 2022 inclusive. The minimum
reported detection magnitude was $m_{min}=21$ (we use `$m$' to
indicate a generic filter magnitude for PS1 which typically uses the
$\rps$ or $\ips$ filters depending on the phase of the moon,
\cite{Schlafly2012}).  We then randomly selected 6 tracklets from the
set of tracklets for each object and randomly selected 4 detections
from tracklets with $>4$ detections.  The $m\ge21$ requirement was
imposed on each detection in an attempt to match the apparent
magnitude distribution of our test data to the apparent magnitude
distribution of F51 observations in the ITF. This is important because
the astrometric uncertainty depends on the apparent magnitude of the
detections and has an impact on the linking and orbit determination
efficiencies.  The algorithm had no difficulty extracting the required
tracklets for 1000 main belt objects but there were only 18 NEOs that
met the requirements.

The time distribution of the randomly selected tracklets
(Figure~\ref{fig:tracklet_time_mag_distn_a}) shows that the number of
tracklets increases with time.  This is due to a major shift in the
F51 system's survey strategy about five years after operations began
and also a secular improvement in the system's capabilities with time.
This distribution should mimic F51's contribution rate to the ITF
under the assumption that the fraction of detected tracklets that are
`isolated' is relatively constant.

The brightest detections in our test data have $m\sim21$ by design
while the faintest objects have reported apparent magnitudes $m>22.5$
(Figure~\ref{fig:tracklet_time_mag_distn_b}). The mode of $m=21.3$
and median of $m=21.4$ of the test data detections are about a half
magnitude brighter than the mode and median of the real F51 ITF
detections of $m=21.9$ and $21.8$, respectively. We were able to
calculate the astrometric error for each detection
(Figure~\ref{fig:tracklet_time_mag_distn_b}) because these objects are
main belt asteroids with precise and accurate orbital elements. The
mean astrometric error is less than 1~arcsec for $m\lesssim22$ and
increases quickly to fainter magnitudes
(Figure~\ref{fig:tracklet_time_mag_distn_b}). As expected, the
astrometric error on these detections is considerably worse than the
mean F51 astrometric error of $\sim 0.17$~arcsec for brighter,
multi-apparition asteroids \citep{Veres2017-astrometricErrors}.


The time between the nearest pairs of tracklets for the same object
has a strong peak at $\ll1$~year because objects are most likely to be
re-detected in the same lunation when they are bright or in a
successive lunation (Figure~\ref{fig:tracklet_time_mag_distn_c}).
Surveys typically re-image the same area of sky even within a lunation
and their field-of-regard is now so large that the same objects can
appear in the data from lunation to lunation. Furthermore, most
objects are brightest and most detectable at perihelion and less
likely to be detected at their next few apparitions. The successive
peaks at multiples of about 1.3~years are simply because the synodic
period of main belt objects is $\sim1.3$~years (for an object with
semi-major axis of $2.5$~au). The time difference between all pairs
of tracklets for the same object also exhibits the 1.3~year synodic
periodicity but the peak at $\ll 1$~year is reduced because the panel
no longer selects the minimum time between pairs of tracklets
(Figure~\ref{fig:tracklet_time_mag_distn_d}).

\subsection{\texttt{Link2} exploration}
\label{s:testlink2}



Recalling that a true linkage includes two tracklets belonging to the
same object, now we define an {\em accurate} linkage as a true linkage
yielding an orbit close to the correct/known one.

To quantify the proximity of two orbits we apply the $D$-criterion
\cite{Dcriterion2000} which measures their distance, $D$, in the space
of the orbital elements $(a,e,i,\Omega,\omega)$. We assume that the
two sets of orbital elements are close enough to consider the linkage
and orbit accurate if $D<0.2$, a commonly used but somewhat arbitrary
value in the literature.
In a case where there are multiple solutions we use the preliminary
solution with the smallest value of the $\chi_4$ norm since, as
discusssed in \cite{Link2Link3}, the values of $D$ and $\chi_4$ are
correlated.

\begin{figure}[!ht]
  \centering
  \includegraphics[width=0.97\textwidth]{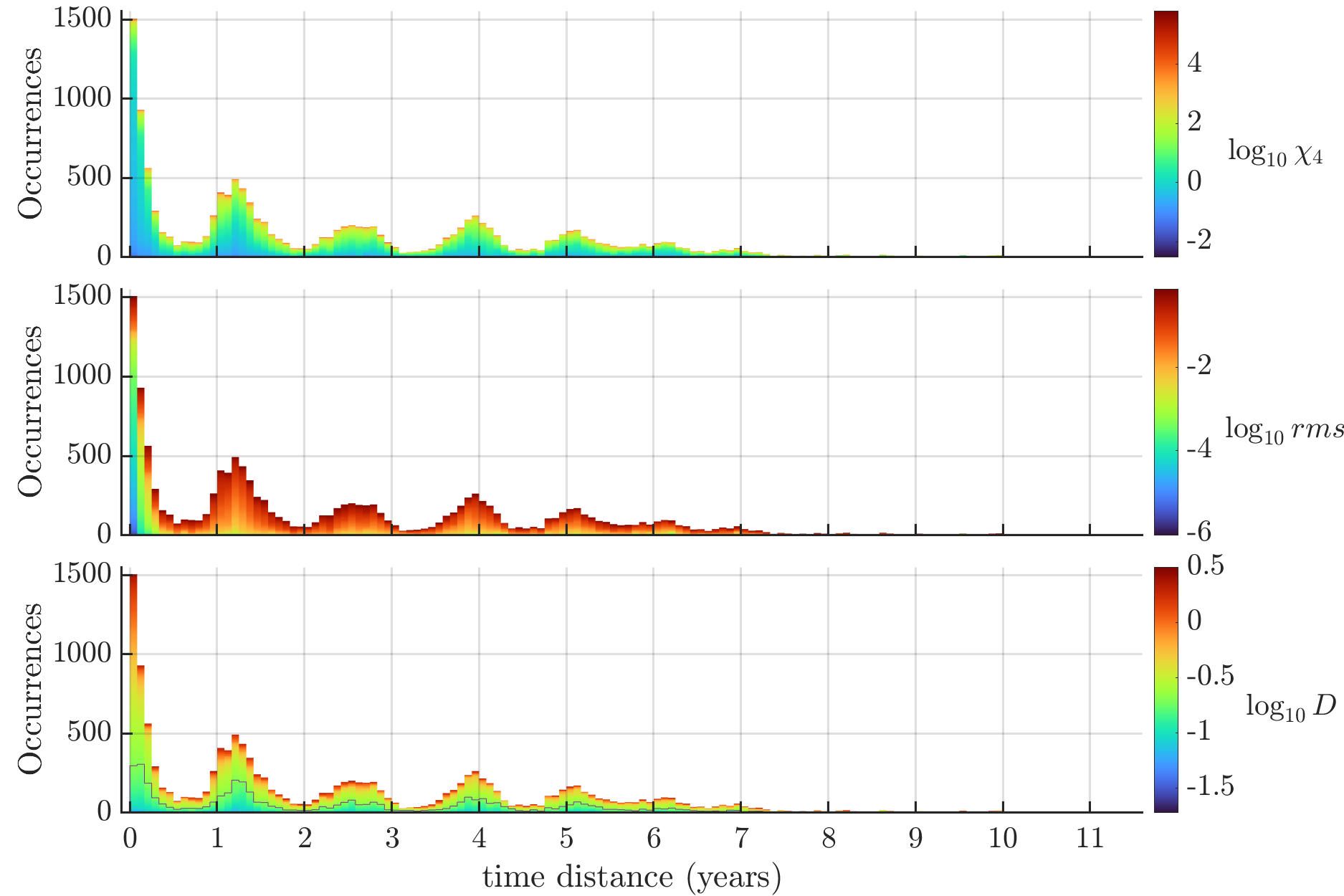}
  \caption{The difference in time between any pair of tracklets
    belonging to the same object. The colors represent the values of
    $\log_{10}\chi_4$ (top), $\log_{10}rms$ (middle), and $\log_{10}D$
    (bottom) of the \texttt{link2} solution with the best value of
    $\chi_4$.}
  \label{fig:timediffcol}
\end{figure}

Our results (Figure~\ref{fig:timediffcol}) obtained by applying
\texttt{link2} to the dataset described in the previous section are
worse than those reported in \cite{Link2Link3}.  Comparing Figures
\ref{fig:tracklet_time_mag_distn_d} and \ref{fig:timediffcol} it is
clear that we miss a considerable number of linkages.
Moreover, the values of the $\chi_4$ and $rms$ indicators are higher
than those in \cite{Link2Link3} due to the fact that here we only
consider observations with an apparent magnitude
$\ge 21$ which have larger astrometric errors than brighter
detections.
In addition, a large number of solutions are lost when the time span
between the mean epochs of the tracklets is too short ($< 14$ days).
Nevertheless, the quality of the preliminary solutions obtained with
\texttt{link2} remains good (Figure~\ref{fig:timediffcol}).

We quantify the \texttt{link2} method's performance for observations
of known objects with respect to the threshold values of $\chi_4$ and
$rms$ with the following metrics:
\[
\begin{aligned}
  \textrm{completeness} &= \cfrac{\#\{\text{true linkages found}\}}{\#\{\mbox{total true linkages}\}}\,,\\
  \textrm{correctness}  &= \cfrac{\#\{\text{true linkages found}\}}{\#\{\text{all linkages found}\}}\,,\\
  \textrm{accuracy}     &= \cfrac{\#\{\text{accurate linkages found}\}}{\#\{\text{true linkages found}\}}\,.
\end{aligned}
\]

\begin{figure}[!ht]
  \centering
  \includegraphics[width=0.32\textwidth]{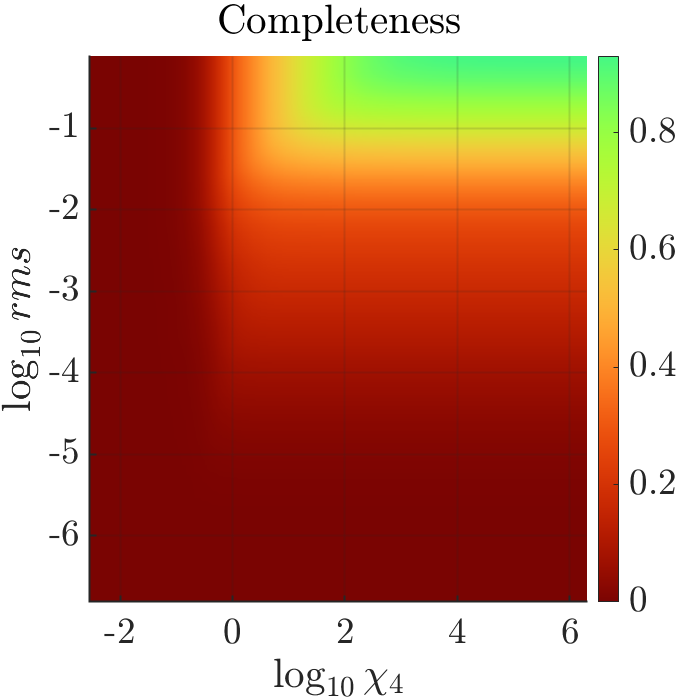}
  \includegraphics[width=0.32\textwidth]{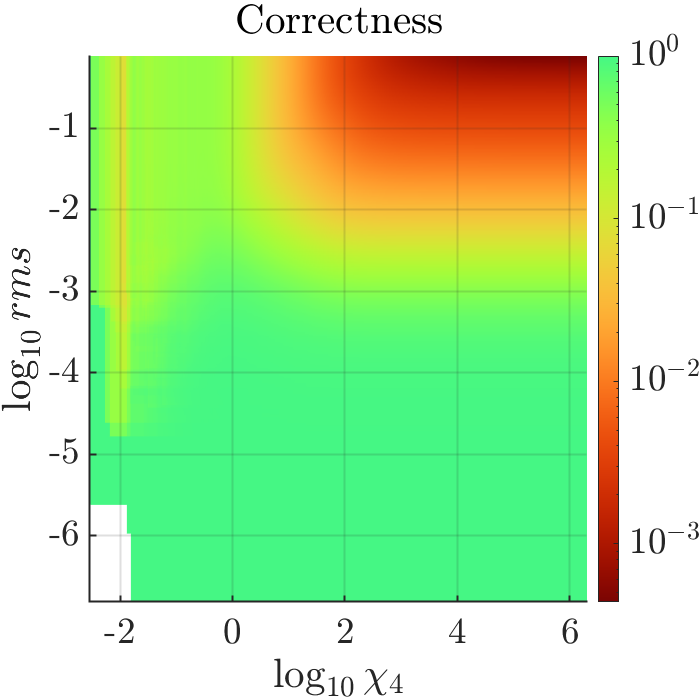}
  \includegraphics[width=0.32\textwidth]{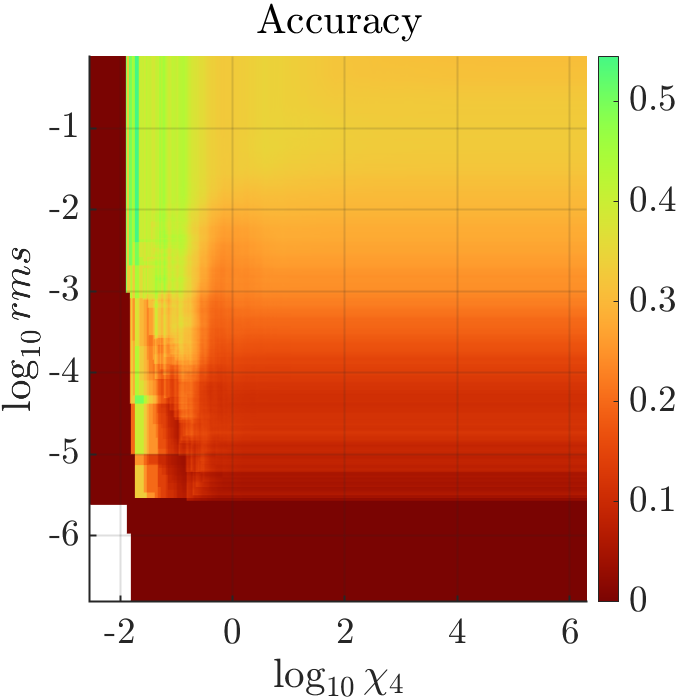}
  \caption{Our algorithm's completeness (left), correctness (middle)
    and accuracy (right) as functions of the $\log_{10}\chi_4$,
    $\log_{10}rms$ thresholds for $D=0.2$.  The white region in the
    middle and right panels corresponds to the cases where no linkages
    were found.}
  \label{fig:complete_correct}
\end{figure}

The completeness, correctness and accuracy
(Figure~\ref{fig:complete_correct}) are consistent with expectations
and with those presented in \cite{Link2Link3} (taking into account the
increased astrometric error in our current set of observations).  A
higher threshold for $\chi_4$ and $rms$ recovers a larger fraction of
possible true linkages at the expense of increasing the number of
false solutions.

\subsection{Constructing LS orbits using 3 tracklets}


\subsubsection{Constructing 3-cycles}

\begin{figure}[!ht]
  \centering
  \includegraphics[width=0.52\textwidth]{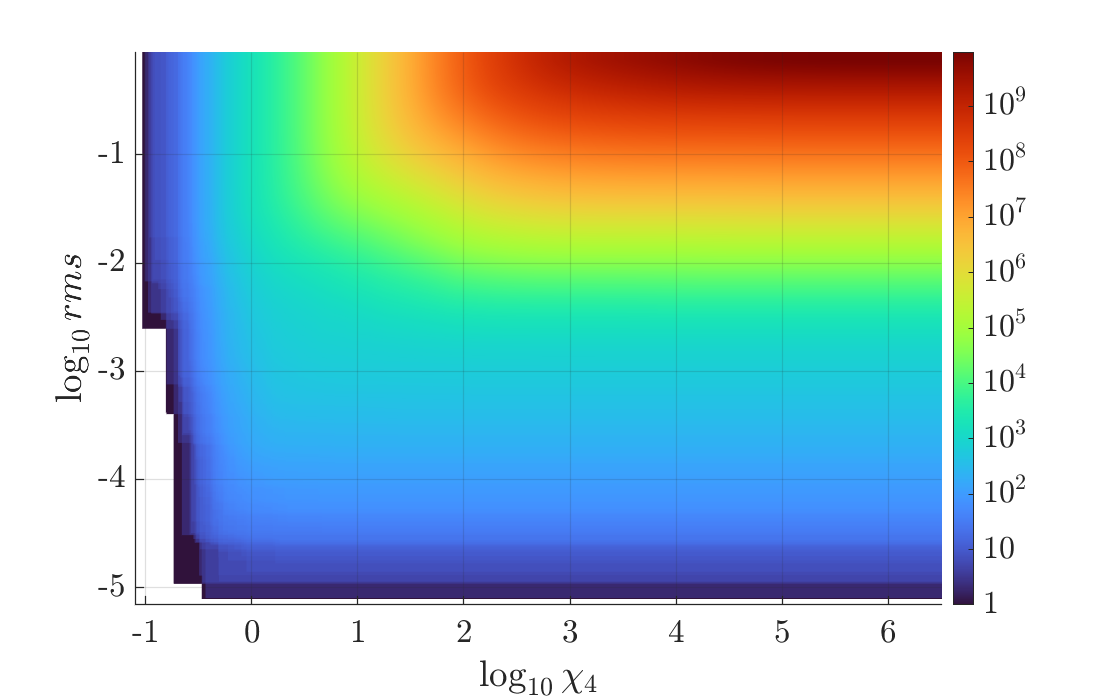}
  \caption{Number of 3-cycles as a function of the maximum allowed
    $\log_{10}\chi_4$ and $\log_{10}rms$.}
  \label{fig:num3cycles}
\end{figure}

The identification of 3-cycles was performed using Algorithm
\ref{al:3cycles}.
Even working with a dataset containing only 6108 tracklets
the total number of 3-cycles would be almost 10 billions if solutions
were not discarded by means of the thresholds on $\chi_4$ and $rms$
(Figure \ref{fig:num3cycles}).

Since our ultimate goal was to apply this procedure to the ITF we
adopted tight thresholds for $\chi_4$ and $rms$ (see the next section)
such that the total number of solutions was manageable.

\subsubsection{Angular momentum norm}

\begin{figure}[!ht]
  \centering
  \includegraphics[width=1\textwidth]{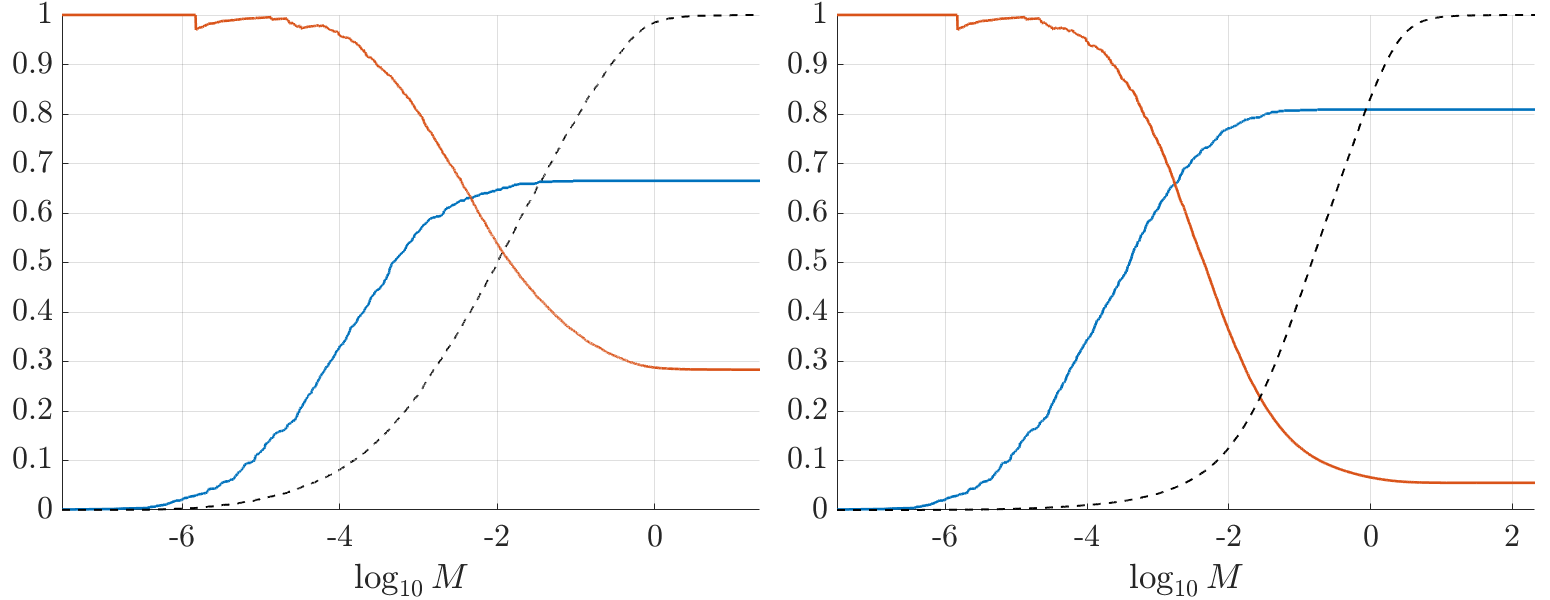}
  \caption{The fraction of objects with at least one true 3-cycle
    (blue), the fraction of true 3-cycles (red), and the fraction of
    accepted 3-cycles (black/dashed) for $\chi_4 = 2$ and $rms =
    1/(10\sqrt{2})$ (left) and $\chi_4 = 5$ and $rms = 0.1$ (right).}
  \label{fig:MhatTest}
\end{figure}

The angular momentum norm (equation \ref{eq:angmomnorm}), hereafter
denoted by $M$, was used to discard {\em false} 3-cycles without
losing too many {\em true} 3-cycles, and the choice of the threshold
value for $M$ depends on the thresholds for $\chi_4$ and $rms$ (Figure
\ref{fig:MhatTest}).

We set the $\chi_4$ and $rms$ thresholds to 5 and 0.1, respectively
(Figure \ref{fig:MhatTest}, right panel), and the threshold value
$\log_{10} M = -1.5$. The values were chosen empirically to produce
manageable results with good efficiency were operationally imposed at
the beginning of the \texttt{link2} exploration (Section
\ref{s:testlink2}).  With these values the procedure identifies at
least one true 3-cycle for more than 80\% of the asteroids and
produces
less than 20,000 3-cycles. Finally, we order the 3-cycles based on
the value of $M$.

\subsubsection{LS orbits}

For each accepted 3-cycle in the previous step we try to construct a
LS orbit from a preliminary orbit computed by either \texttt{link2} or
\texttt{link3} (see Section~\ref{s:LSorb}) using all the observations
within the 3-cycle's. Most ($>92$\%) of the false 3-cycles do not
converge to a LS orbit but the majority ($\sim92$\%) of the true
3-cycles do converge.


\begin{figure}[!ht]
  \centering
  \includegraphics[width=0.49\textwidth]{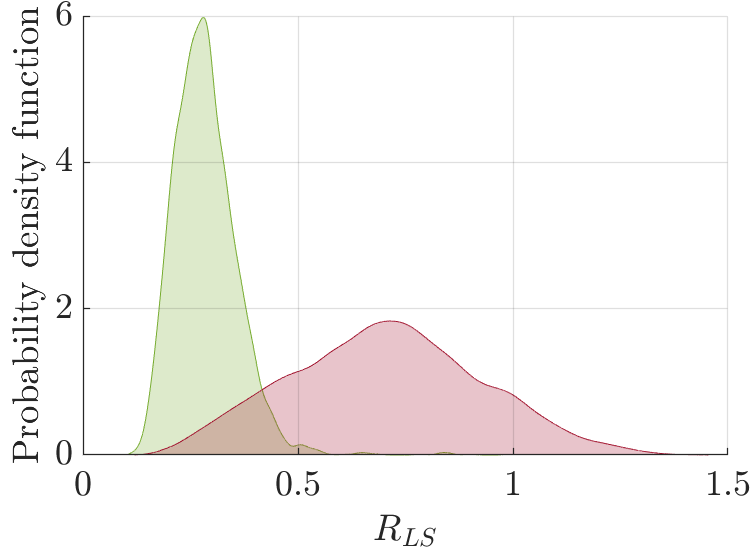}
  \caption{PDF of the astrometric residuals, $R_{LS}$, for the false
    (red) and true (green) LS orbits computed from the 3-cycles.}
  \label{fig:3cyc}
\end{figure}

The quality of the LS orbits is assessed by the rms of the residuals,
$R_{LS}$, which is used to discard most of the false 3-cycles (Figure
\ref{fig:3cyc}).
The maximums of the PDFs for the true and false LS orbits are
well-separated but the tail of the false LS distribution overlaps
almost completely with the $R_{LS}$ for the true LS orbits.  We
selected a threshold value of $R_{LS} \leq 0.5$ to accept $\approx
99$\% of true LS orbits at the cost of also accepting $\approx19$\% of
false LS orbits. The remaining false orbits are mostly eliminated in
the next step (\S\ref{ss.Joining4OrMorTracklets}) by searching for
additional isolated tracklets that are consistent with each orbit.

\begin{figure}[!ht]
  \centering
  \includegraphics[width=0.49\textwidth]{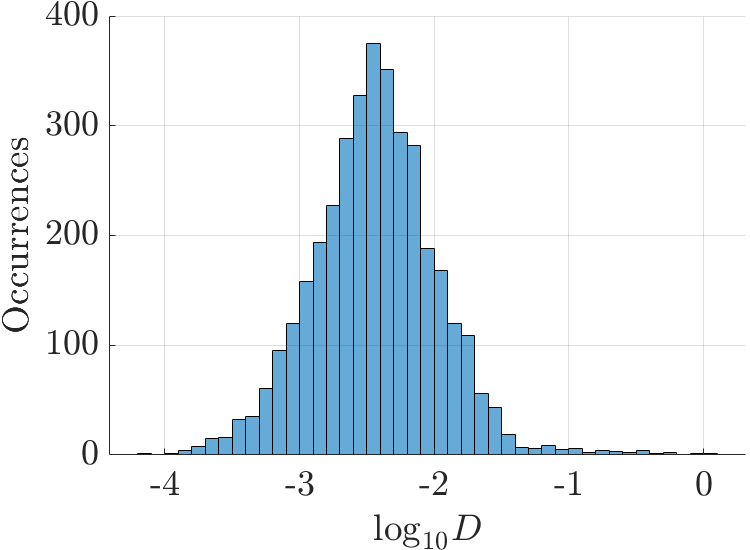}
  \caption{Distribution of the values of $D$ for the accepted LS
    orbits computed from the 3-cycles.}
  \label{fig:3cycD}
\end{figure}


After applying the $R_{LS}$ we find at least one LS orbit for
$\sim78$\% of the asteroids and the quality of these orbits is good as
demonstrated with the $D$-criterion of our LS orbit compared to the
known, high-accuracy orbit (Figure \ref{fig:3cycD}). We find that
$\approx 99.6 \%$ of the orbits have $D<0.2$, the value we used above
to determine if two sets of orbital elements were similar.

\subsection{Joining 4 or more tracklets}
\label{ss.Joining4OrMorTracklets}

The final step is to apply Algorithm 2 to identify more tracklets and
use all the detections contained in the tracklets to calculate an LS
orbit. The LS orbit is accepted if $R_{LS}<0.5$.

\begin{figure}[!ht]
  \centering
  \includegraphics[width=0.49\textwidth]{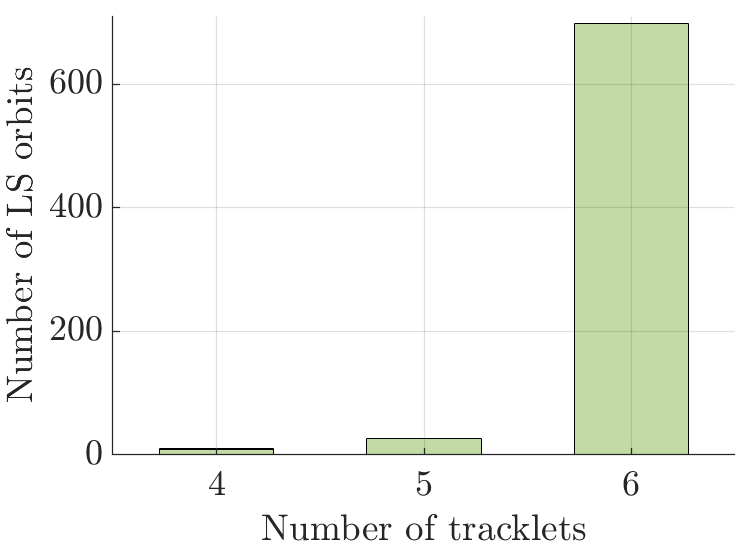}
  \includegraphics[width=0.49\textwidth]{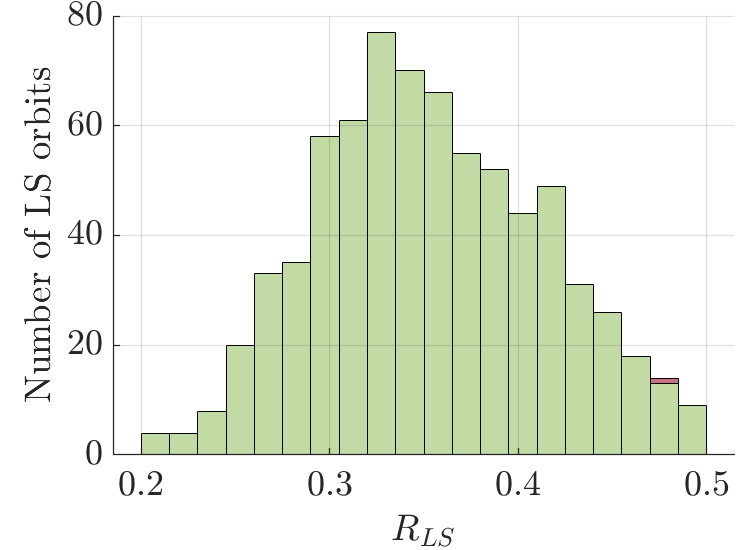}
  \caption{(Left) The number of accepted LS orbits computed from 4, 5,
    and 6 tracklets. (Right) The values of $R_{LS}$ for the accepted
    orbits where green entries represent correct LS orbits and the
    single red entry indicates the single false orbit.}
  \label{fig:LS}
\end{figure}

The algorithm yielded 735 accepted LS orbits with 4 or more tracklets
of the test sample (Figure~\ref{fig:LS}, left) and only one of them is
false. The single incorrect orbit includes 3 tracklets from one
asteroid and 1 tracklet of another object.  A total of 698 true orbits
included all 6 possible tracklets in the test sample for each
asteroid.

\begin{figure}[!ht]
  \centering
  \includegraphics[width=0.49\textwidth]{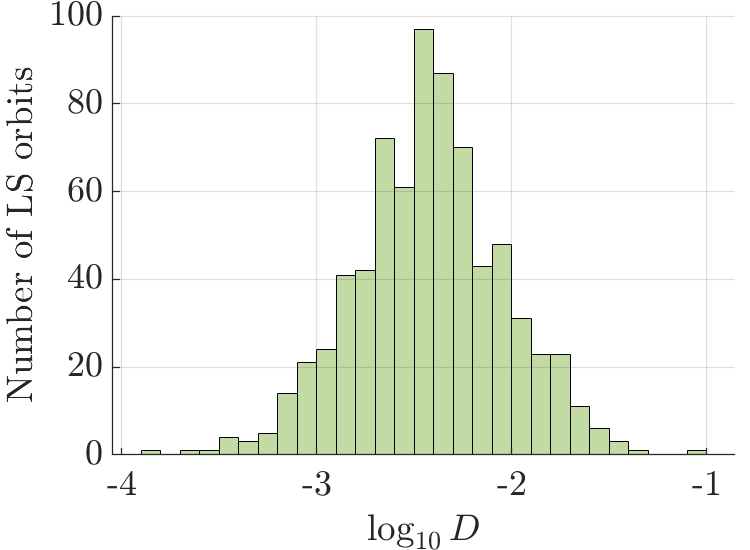}
  \caption{Values of the $D$-criterion for the final accepted LS
    orbits.
  }
  \label{fig:LSD}
\end{figure}

The average astrometric rms of the accepted orbits is
$\approx0.35$~arcsec (Figure~\ref{fig:LS}, right), much better than
the mean error of the detections in our test sample
(Figure~\ref{fig:tracklet_time_mag_distn}b).
Similarly, the average $D$-criterion for the accepted orbits compared
to the actual orbits is $\sim0.0056$, significantly better than the
maximum value of $0.2$ used as a threshold when setting our metric
thresholds (Figure~\ref{fig:LSD}).



The efficiency for recovering NEOs is $72^{+9}_{-11}$\% consistent
with the $72.1\pm0.1$\% MBA detection efficiency.  The average
$D$-criterion for the NEOs is $0.0026$.


\section{Application to the ITF}
\label{s:itfproc}

We applied the procedure outlined in the preceding sections to the
Pan-STARRS observations in the ITF as of 2022 July 30, after the work
of \cite{HelioLinC2018} and \cite{WerykITF2020}, a dataset containing
3,760,777 F51 observations.

We first applied corrections for two types of inconsistencies in the
data: 1) duplicate observations with the same RA and declination but
at slightly different epochs and 2) tracklets spanning too long a time
range.

There were only about half a dozen duplicate observations that had
identical values of RA and declination at two or more times that
differed by only a few seconds.  Duplicate observations were combined
into a single detection with the same RA and declination at the
average time of observation.


Tracklets are generally a set of observations acquired over a short
period of time within a single night so we split a tracklet into
sub-tracklets if the time separation between two consecutive
observations was $>0.5$~days. In one extreme case, a single ITF
tracklet contained observations spanning from 2014 June 21 to 2014
September 12.

\begin{figure}[!ht]
  \centering
  \includegraphics[width=0.47\textwidth]{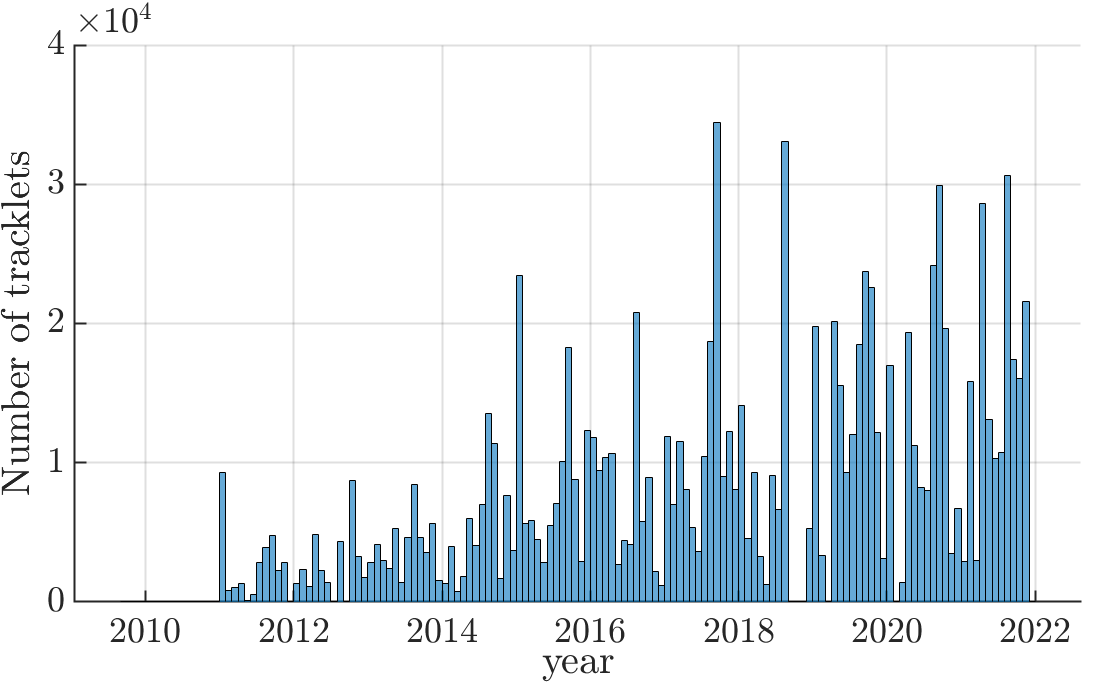}
  \includegraphics[width=0.47\textwidth]{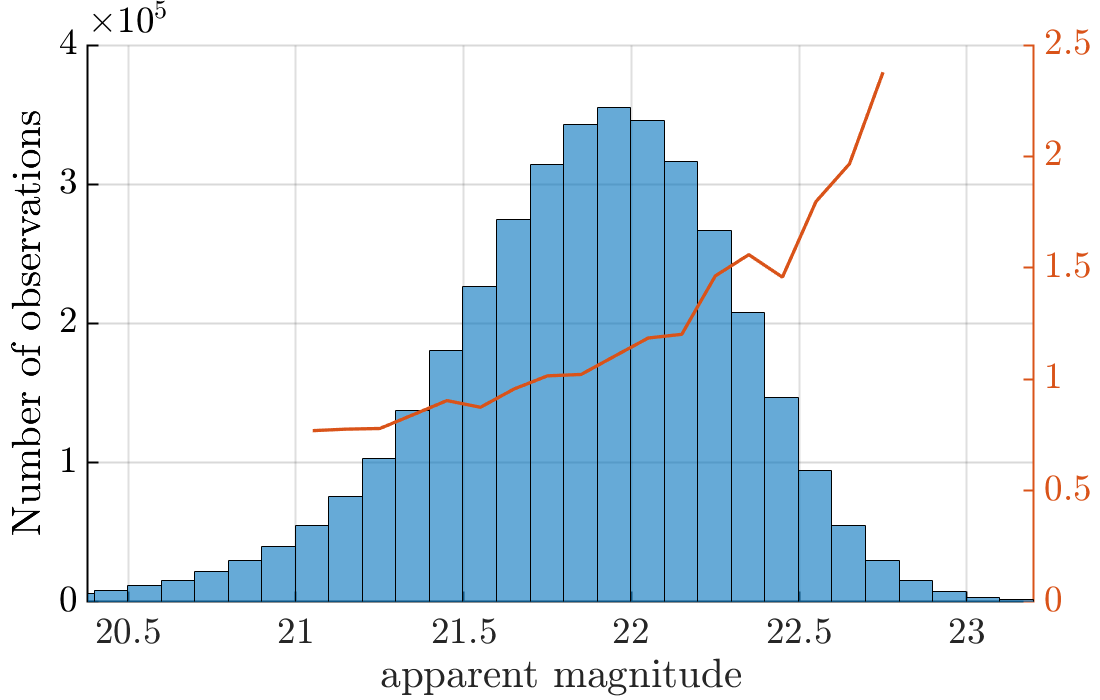}
  \caption{
    (left) The times of observation of F51 tracklets in the ITF and
    (right) their reported apparent magnitude distribution and (in
    red) their average astrometric uncertainty as a function of the
    reported apparent magnitude as calculated from the 1000 object
    test sample.}
  \label{fig:tracklet_time_mag_distn_ITF}
\end{figure}

After applying these cleaning operations and only selecting tracklets
with at least 3 observations we were left with 3,693,929 detections
contained in 1,072,171 tracklets. The distribution of the times of
observations (Figure~\ref{fig:tracklet_time_mag_distn_ITF}, left)
reflects the operations of the Pan-STARRS survey which began science
operations in 2010 \cite{chambers2019panstarrs1} with an increasing
fraction of time devoted to asteroid surveying as the years passed and
gradual improvement in the system's asteroid detection efficiency.

The apparent magnitudes of the F51 ITF observations
(Figure~\ref{fig:tracklet_time_mag_distn_ITF}) are typically greater
than $21.7$, the system's limiting magnitude in their most sensitive
wide-band filter, $w_{\rm P1}$, that was used for most asteroid
surveying \cite{Denneau2013-MOPS}. With $\gtrsim 59$\% of the
observations greater than the system's limiting magnitude the
astrometric uncertainty on these observations is much worse than the
system's average rms uncertainty of $\sim0.13$" on observations of
brighter targets \cite{Milani2012-Identification}. The mode of the
astrometric uncertainty is almost $8\times$ higher at about 1" and
larger than 2" for the faintest reported ITF observations
(Figure~\ref{fig:tracklet_time_mag_distn_ITF}, right).

\begin{figure}[!ht]
  \centering
  \includegraphics[width=0.48\textwidth]{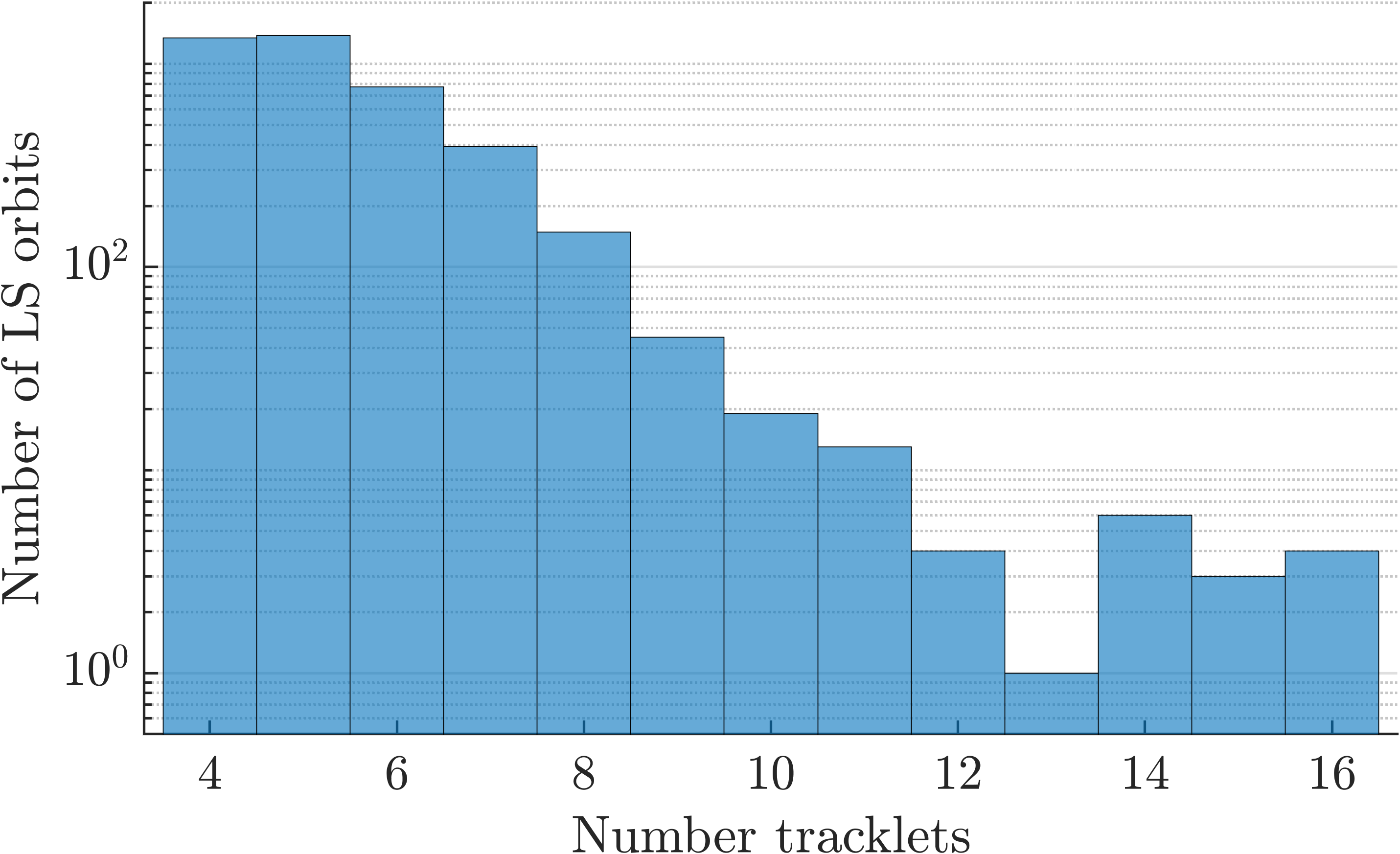}
  \includegraphics[width=0.48\textwidth]{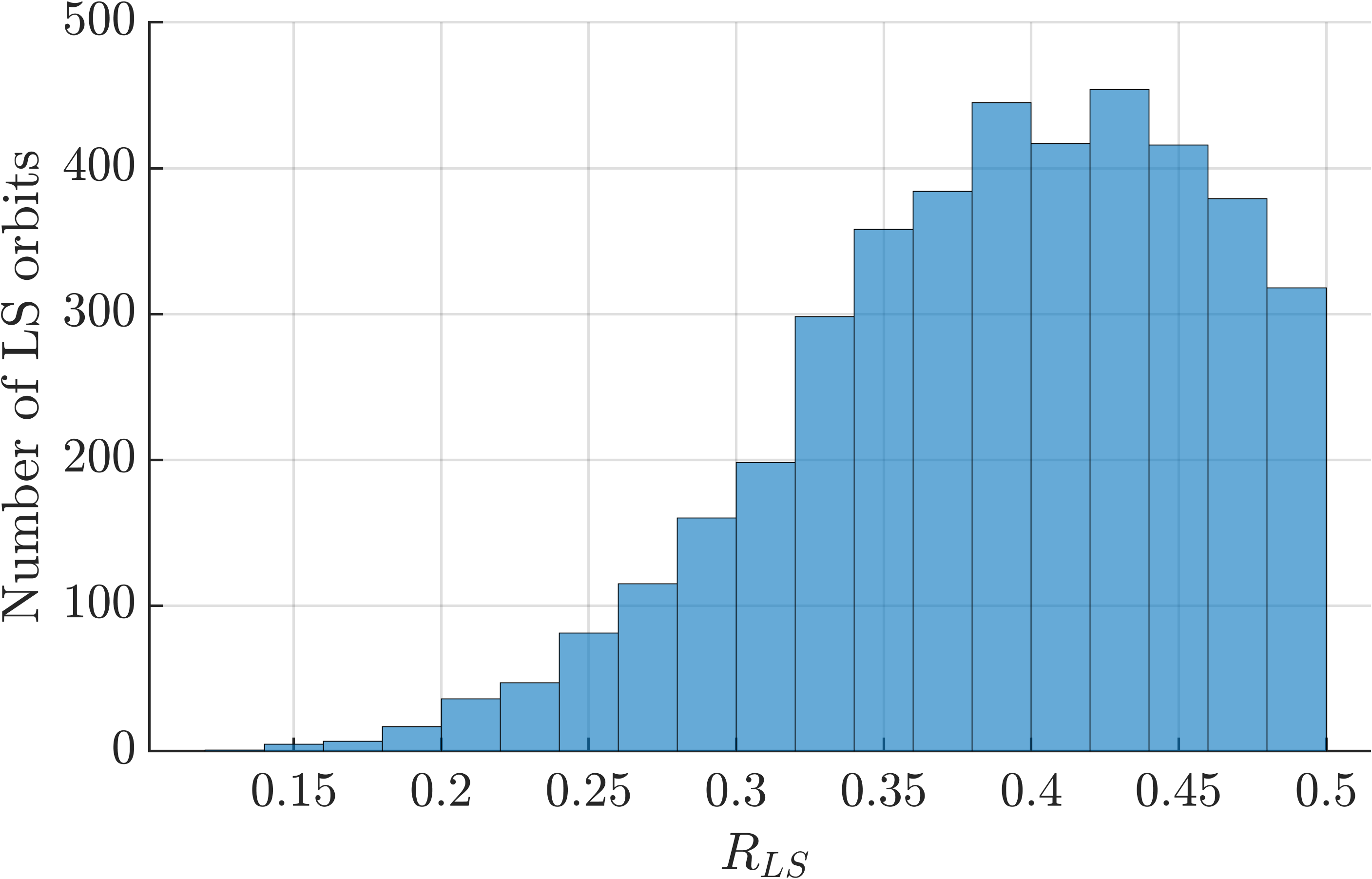}
  \caption{
    (left) The number of tracklets per object in the final set of LS
    orbits.    
    (right) The values of the residuals of the final set of LS
    orbits.}
  \label{fig:numTraOrbITF}
\end{figure}

\begin{figure}[!ht]
  \centering
  \includegraphics[width=0.48\textwidth]{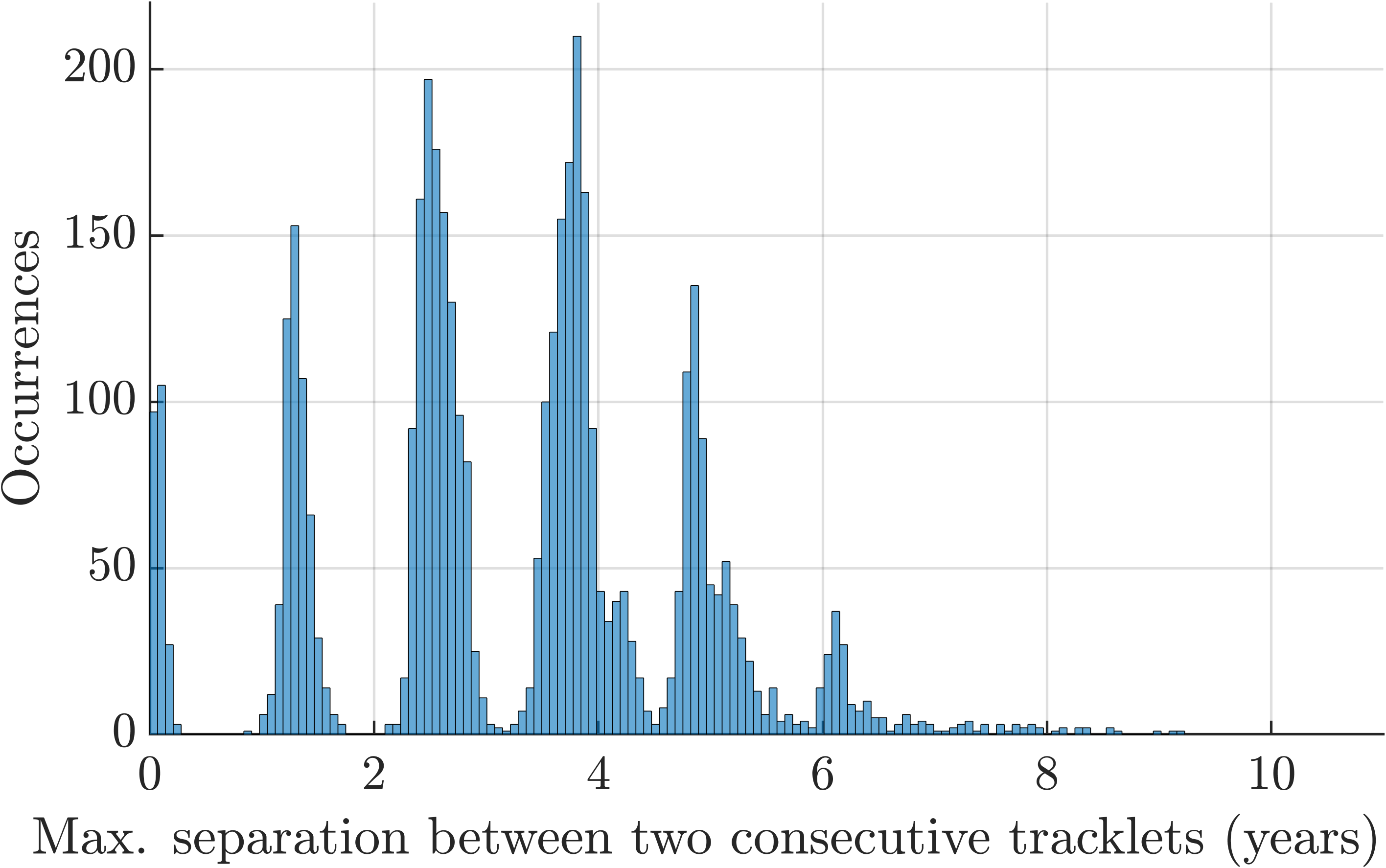}
  \includegraphics[width=0.48\textwidth]{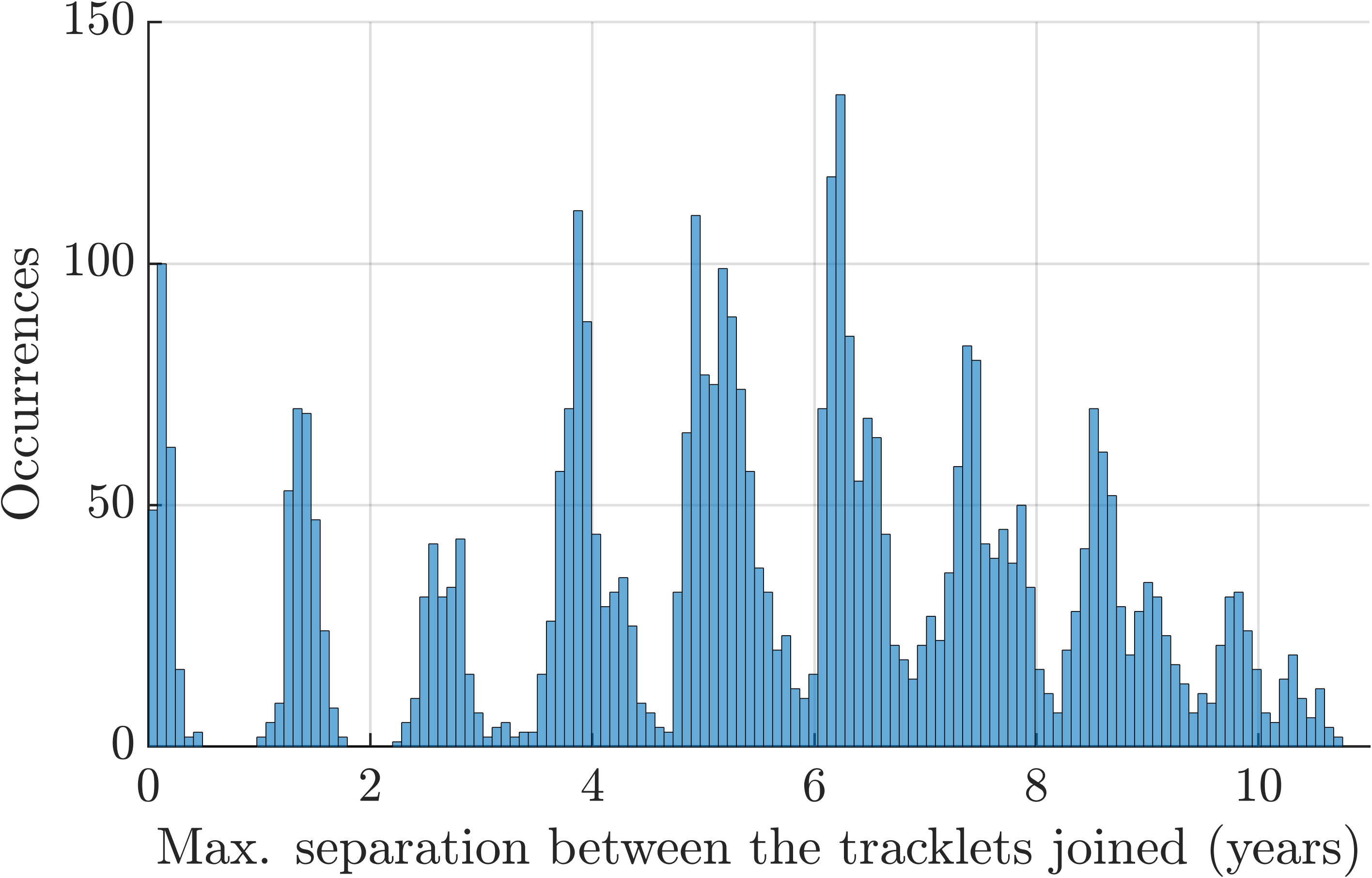}
  \caption{   
    (left) The maximum time between any two sequential tracklets
    within the set of tracklets belonging to a single orbit.   
    (right) The total time span of the tracklets in the final set of
    orbits.}
  \label{fig:distTrmsITF}
\end{figure}

The procedure described in this paper was applied to the cleaned F51
ITF observations and identified
4,135 LS orbits that included 4 or more tracklets in 4 different
nights. The vast majority of the orbits contain only 4 tracklets but
four of the orbits contain 16 tracklets
(Figure~\ref{fig:numTraOrbITF}). The maximum time separation between
two sequential tracklets in a single orbital solution spans a wide
range (Figure~\ref{fig:distTrmsITF}, left) from less than one year to
almost 10 years, with peaks corresponding to the synodic periods of
main belt objects. The total time span of the observations linked to
a single object exhibits similar peaks with a maximum greater than 10
years and a mean greater than 5 years (Figure~\ref{fig:distTrmsITF},
right).

\begin{figure}[!ht]
  \centering
  \includegraphics[width=0.67\textwidth]{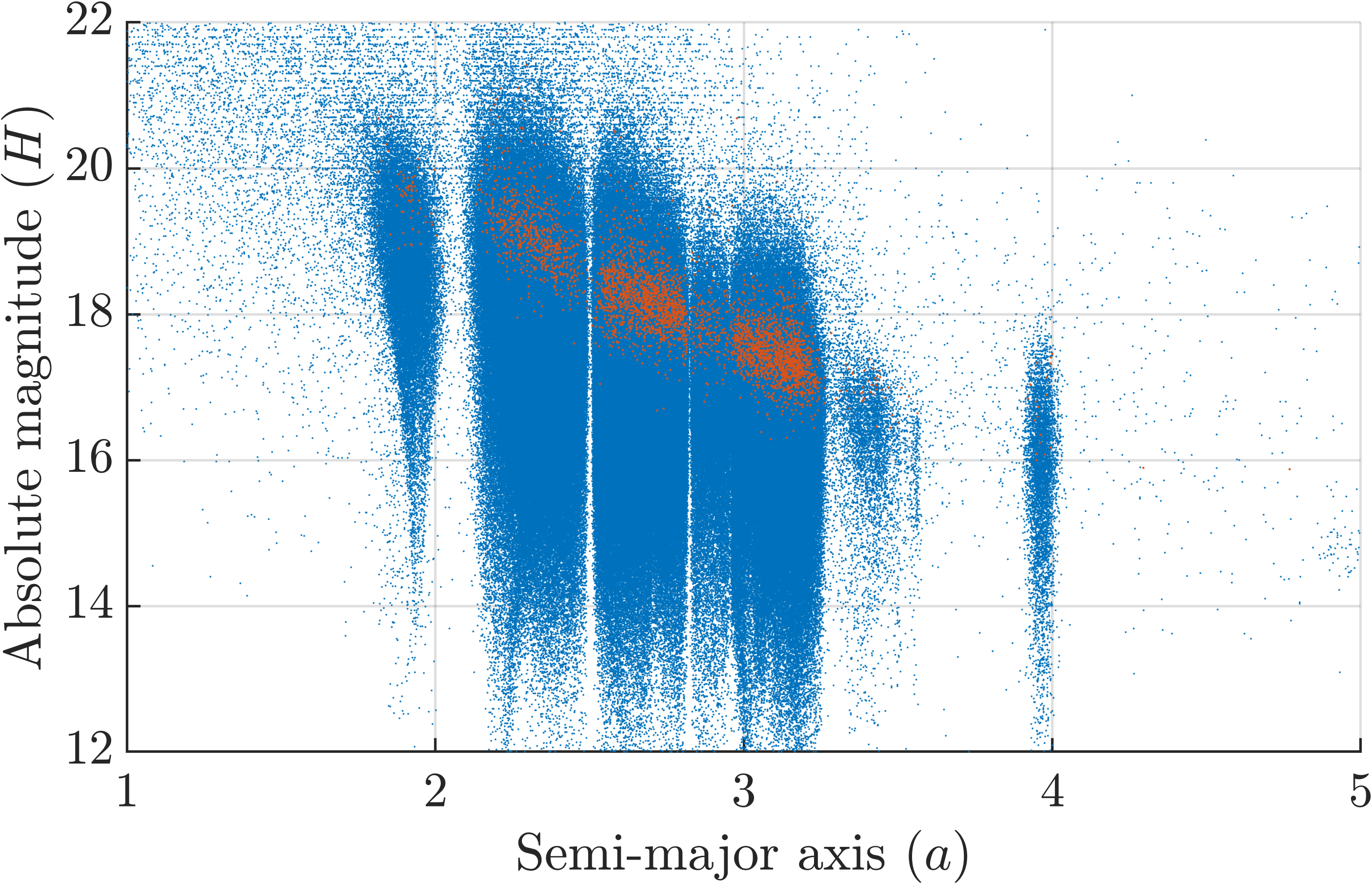}
  \caption{Absolute magnitude versus semi-major axis of (blue) known
    objects from JPL Horizons$^2$ and (orange) this work.}
  \label{fig:Ha}
\end{figure}

\begin{figure}[!ht]
  \centering
  \includegraphics[width=0.98\textwidth]{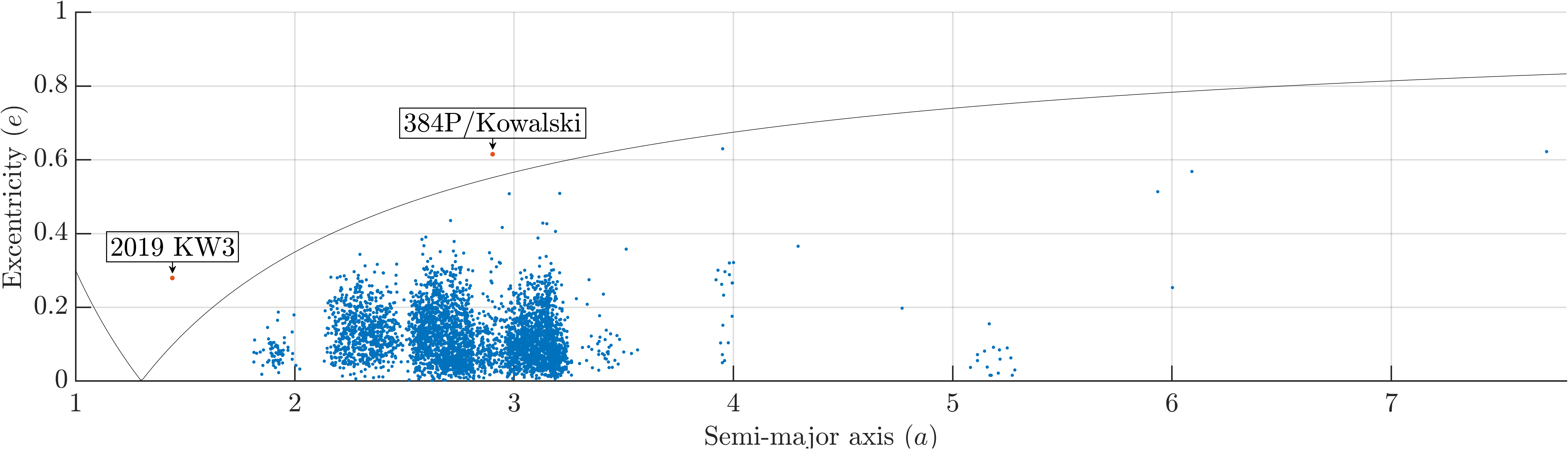}
  \includegraphics[width=0.98\textwidth]{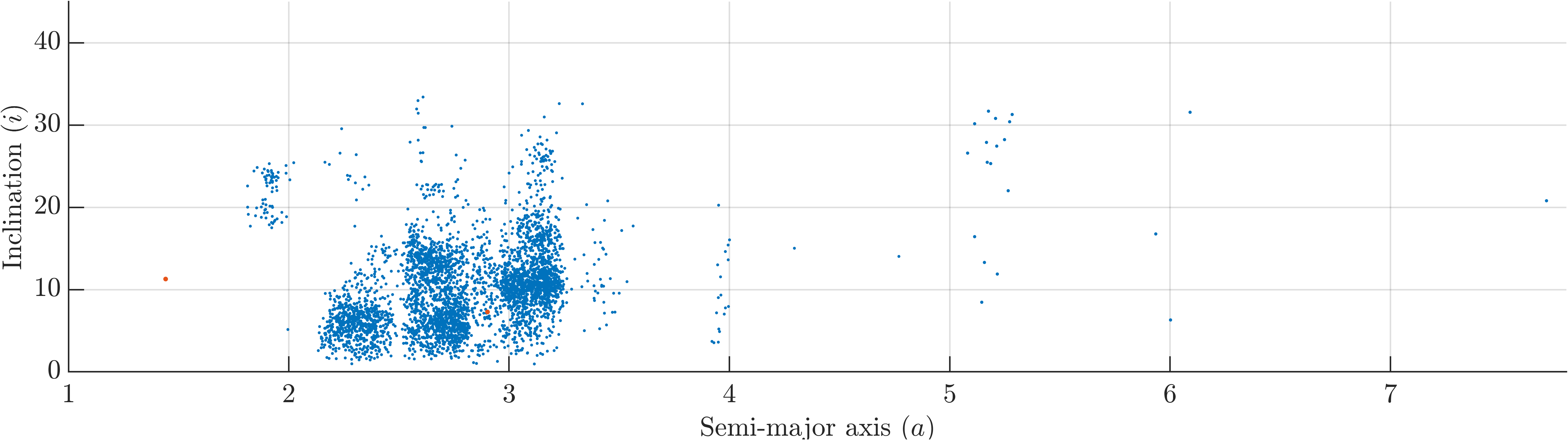}
  \caption{Eccentricities (top) and inclinations (bottom, in degrees)
    versus semi-major axis of the accepted LS orbits. The red dots
    above the curve on the top panel correspond to NEOs.}
  \label{fig:aeiITF}
\end{figure}

The LS orbits have much higher astrometric residuals than typical of
F51 because they only include detections at much fainter magnitudes
(Figure~\ref{fig:Ha}),
with the peak of the distribution at $\sim0.45$", almost 50\% higher
than the test dataset's residuals that were specifically designed to
match the apparent magnitudes of F51's ITF detections
(Figure~\ref{fig:LS}). We think that the high residuals are not due
to the presence of false solutions because 1) we only found one false
solution out of 728 in our test dataset and 2) the orbital
distribution of the final set of orbits is a good match to the orbital
distribution of known objects as we will show below
(Figure~\ref{fig:aeiITF}). Future implementations of our algorithm
should consider relaxing the constraint on $R_{LS}$ to identify more
LS orbits consistent with the observations's astrometric uncertainties
and adding constraints on the number of tracklets/nights included in
the orbit.

There is evidence to support the conclusion that the 4,135 LS orbits
correspond to real objects because their orbit distribution reproduces
the distribution of objects in the main belt including revealing
Kirkwood Gaps, Jupiter Trojans, and both collisional and dynamical
asteroid families within the main belt (Figure~\ref{fig:aeiITF}).
Almost all the orbits correspond to MBAs but 2 represent NEOs
(384P/Kowalski and 2019 KW3) and the most distant object with a
semi-major axis of $\sim7.7$~au corresponds to a Centaur.









The absolute magnitudes of our linkages also provide evidence that
they are legitimate (Figure~\ref{fig:Ha}). They are strongly skewed
to the faint end of the main belt values because objects with smaller
$H$ are more likely to have been detected often and objects with
larger $H$ are likely too faint to be detected regularly.  i.e. The
smallest objects might be detected in a serendipitous apparition when
they are at perihelion near opposition but are unlikely to be detected
in subsequent apparitions.  More than 99\% of the asteroids we
identified in the inner belt ($a<2.5\au$) are sub-km diameter
asteroids with $H>17.6$ assuming an S-class albedo of $0.17$ typical
of objects in the main belt
\cite{DeMeo2015-MBstructure,Wright2016-WISEAlbedos}. In 2009 it was
suggested \cite{Gladman2009} that the main belt population
($2.0\au<a<3.5\au$) is completely known for $H<15$ and only
$\sim0.02$\% of the objects we identified fall into that absolute
magnitude range, i.e. 2 objects, both in the outer region of the belt.
About 25\% of our main belt objects have $H<17.5$, the completeness
limit proposed in 2015 \cite{Denneau2015-catastrophicDisruptionRates},
inhabiting the outer regions of the belt.  Given that the outer belt
is dominated by low albedo (typically $\sim0.03$
\cite{DeMeo2015-MBstructure,Wright2016-WISEAlbedos}) C-class
asteroids, a $1\km$ diameter asteroid in the outer belt would have
$H\sim19.4$, suggesting that it will take some time till the main belt
is effectively complete for km-scale asteroids.

\begin{figure}[!ht]
  \centering
  \includegraphics[width=0.67\textwidth]{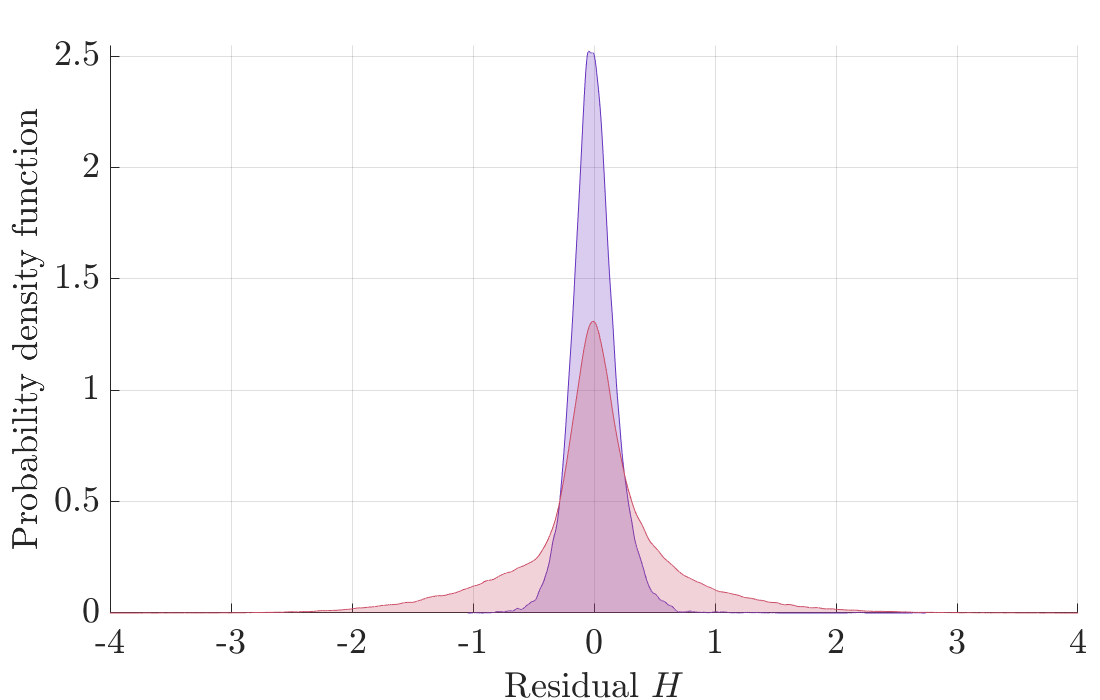}
  \caption{Probability density function of the $H$ residuals for
    (blue) the LS orbits obtained with the test data and (red) the LS
    orbits identified in the ITF data.}
  \label{fig:Hres}
\end{figure}

The half-width at half-maximum (HWHM) of the distribution of $H$
residuals for the main belt test data (Figure~\ref{fig:Hres}) is about
0.2~mags implying an SNR$\,\sim5$ for the detections in the test data.
This is about what is expected for objects with the magnitude
distribution having a mode of $V\sim21.3$
(Figure~\ref{fig:tracklet_time_mag_distn}), almost half a magnitude
brighter than the system limiting magnitude where each detection
typically has SNR$\,\sim3$.  The HWHM of the distribution of the ITF
objects with LS orbits is only 25\% larger at $\sim0.25$~mags but this
comparison does not capture the different shapes of the two
distributions.  The $H$ residuals for the ITF orbits have a much wider
range of values extending out to $\sim2$~mags due to the detections
being typically fainter than the system limiting magnitude.

The MPC has a strict set of criteria for submitting candidate linkages
of objects in the ITF that are intended to reduce the likelihood that
the linkages are false. Basically, they require multiple tracklets
within the same apparition and only a small fraction of the 4,135
objects that passed our LS orbit procedure met the MPC's submission
criteria and only 112 had not already been identified.  Of those 112
candidates, 107 were accepted by the MPC as new designations, i.e. new
discoveries, 4 were identified as known objects, and the last object
was a new discovery but included two mis-identified tracklets.

%

\section{Conclusions}

We presented a procedure to join tracklets in large datasets based on
the Keplerian integrals method {\tt link2} which allowed us to link
tracklets that may be separated by years-long gaps in time. The
quality of the accepted solutions are assessed by different norms to
ensure that the final results are reliable. The procedure is fast
enough that a complete exploration of a large dataset is
computationally feasible and it was applied to F51 observations in the
ITF yielding more than 4,000 orbits, mostly MBAs, but also 2 NEOs.

Despite the success of our method $<1$\% of our recovered orbits meet
the MPC's current requirements for submission of new orbit
identifications.
The MPC requires that an ITF identification contains at least 4
tracklets acquired over a minimum of 4 separate nights with an
observational arc spanning at least 10 days if all tracklets pertain
to a single apparition. For identifications over multiple apparitions
the MPC requires that at least one of the apparitions contains at
least 3 tracklets obtained over a minimum of 3 separate nights and the
other apparitions must contain at least 2 tracklets acquired on at
least 2 nights. These criteria make perfect sense for classical
methods, as most of them require that the time separation between the
tracklets is not too large to compute a preliminary orbit, but could
be relaxed with the implementation of new methods that take into
account additional constraints, like the algorithm presented in this
work.
\section*{Acknowledgments}

This work was supported by the Spanish State Research Agency through
the Severo Ochoa and María de Maeztu Program for Centers and Units of
Excellence in R\&D (CEX2020-001084-M) and through the H2020 MSCA ETN
Stardust-Reloaded, Grant Agreement Number 813644.  Ó. Rodríguez was
supported by the Spanish MINECO/FEDER grant PID2021-123968NB-I00
(AEI/FEDER/UE). G. F. Gronchi and G. Ba\`u acknowledge the Italian
project MIUR-PRIN 20178CJA2B ``New frontiers of Celestial Mechanics:
theory and applications'' and the GNFM-INdAM.

\section*{Data Availability}

The data underlying this article will be shared on reasonable request
to the corresponding author.

\printbibliography

\end{document}